%% file: GASS3505_accepted.tex
\documentclass[useAMS,usenatbib]{mn2e}
\usepackage{setspace}

\usepackage{caption}
\usepackage[english]{babel}
\usepackage{graphicx}
\usepackage{subfigure}

\usepackage{rotating}
\usepackage{stfloats}
\usepackage{txfonts} 
\usepackage{textcomp}	
\usepackage{natbib}
\usepackage{amssymb}
\usepackage{graphicx}
\usepackage{tabularx}
\usepackage[english]{babel}
\usepackage[T1]{fontenc}
\usepackage{multirow}

\usepackage{threeparttable}

\def\HI{H{\,\small I}}

\newcommand{\kms}{$\,$km$\,$s$^{-1}$}

\newcommand{\msun}{{${\rm M}_\odot$}}

\def\HI{H{\,\small I}}

\def\emph#1{{\sl #1}}
\newcommand{\ltsima} {$\; \buildrel < \over \sim \;$}
\newcommand{\gtsima} {$\; \buildrel > \over \sim \;$}
\newcommand{\lta} {\lower.5ex\hbox{\ltsima}}
\newcommand{\gta} {\lower.5ex\hbox{\gtsima}}

\newcommand{\atlas}{{ATLAS$^{\rm 3D}$}} 
\newcommand{\mhi}{{M$_{\rm HI}$}}

\title{GASS 3505: the prototype of HI-excess, passive galaxies}
\author[K. Ger\'{e}b et al.]
{K. Ger\'{e}b$^1$\thanks{E-mail:
kgereb@swin.edu.au},
B. Catinella$^{2}$,
L. Cortese$^{2}$,
K. Bekki$^{2}$,
S. Moran$^{3}$,
D. Schiminovich$^{4}$
\\
$^1$Centre for Astrophysics and Supercomputing, Swinburne University of Technology, Hawthorn, VIC 3122, Australia\\
$^2$ICRAR, M468, The University of Western Australia, 35 Stirling Highway, Crawley, Western Australia, 6009, Australia \\
$^3$Harvard-Smithsonian Center for Astrophysics, 60 Garden St, Cambridge, MA 02138, USA \\
$^4$Department of Astronomy, Columbia University, 550 West 120th Street, New York, New York 10027, USA}

\begin{document}


\pagerange{\pageref{firstpage}--\pageref{lastpage}} \pubyear{2015}

\maketitle

\label{firstpage}

\begin{abstract}
We present our multiwavelength analysis of a prototype \HI-excess galaxy, GASS 3505, selected based on having a large gas content ($M_{\rm HI} = 10^{9.9}$ \msun) compared to its little associated star formation activity ($\sim$0.1 \msun\ yr$^{-1}$) in the GALEX Arecibo SDSS Survey (GASS). 
Very Large Array (VLA) observations show that the \HI\ in GASS 3505 is distributed in a regularly rotating, extended ($\sim$50 kpc radius) gas ring. In the SDSS optical image GASS 3505 appears as a bulge-dominated galaxy, however deep optical imaging reveals low surface brightness ($\gtrsim25$ mag arcsec$^{-2}$) stellar emission around the central bulge.
Direct evidence for accretion is detected in form of an extended ($\sim$60 kpc) stellar stream, showing that GASS 3505 has experienced a minor merger in the recent past. We investigate the possibility that the \HI\ ring in GASS 3505 was accreted in such a merger event using N-body and smoothed particle hydrodynamic (SPH) simulations. The best model that reproduces the general properties (i.e., gas distribution and kinematics, stellar morphology) of the galaxy involves a merger between the central bulge and a gas-rich ($M_{\star}$ = 10$^9$ \msun\ and $M_{\rm HI}$/$M_{\star}$ = 10) disk galaxy. However, small discrepancies in the observed and modeled properties could suggest that other sources of gas have to be involved in the build-up of the gas reservoir. This work is the first step toward a larger program to investigate the physical mechanisms that drive the large scatter in the gas scaling relations of nearby galaxies.

\end{abstract}


\input{Tex_files/Intro.tex}


\input{Tex_files/GASS3505_Selection.tex}


\input{Tex_files/HI_obs.tex}


\input{Tex_files/Optical_data.tex}


\input{Tex_files/UV_data.tex}


\input{Tex_files/K_S.tex}


\input{Tex_files/Environment.tex}


\input{Tex_files/Context_otherSurveys.tex}


\input{Tex_files/Simulation.tex}


\input{Tex_files/Discussion.tex}


\section{Acknowledgements}
We thank the anonymous referee for the useful comments that helped us to improve the manuscript.
KG thanks Tobias Brown, Virginia Kilborn, and Katharina Lutz for helpful discussions on this project.
BC is the recipient of an Australian Research Council Future Fellowship (FT120100660). 
BC,KB and LC acknowledge support from the Australian Research Council's Discovery Projects funding scheme (DP130100664 and DP150101734). BC thanks Jacqueline van Gorkom and Baerbel Koribalski for help with the preliminary reduction of VLA data used in this paper.
The National Radio Astronomy Observatory is a facility of the National Science Foundation operated under cooperative agreement by Associated Universities, Inc.

\appendix 
\input{Tex_files/Appendix.tex}

\end{document}

%% file: Tex_files/Intro.tex
\section{Introduction}

Cold gas provides the reservoir for star formation in galaxies from the distant to the nearby Universe. Our understanding of the mode and rate of accretion onto galaxies is of crucial importance, because fresh supplies of gas are needed for fuelling ongoing star-forming (SF) processes in galaxies \citep{Larson1972}. 

Simulations predict that cosmological gas accretion onto galaxies happens in cold or hot mode, depending on the environment and redshift \citep{Binney2004, Keres2005, Dekel2009}. Accretion along cosmic filaments has been probed at intermediate redshift ($0.3 < z < 1$) using low-ionization absorption line observations, such as MgII absorbers against strong background quasars (see review by \citealt{{Churchill2005}}). These studies have demonstrated that MgII absorption systems are likely falling in towards galaxies via cold flows \citep{Kacprzak2012, Martin2012}. In the nearby Universe it is expected that gas-rich satellites and gas infall from the intergalactic medium are the main contributors to accretion processes and subsequent galaxy growth \citep{Sancisi2008}. Neutral hydrogen (\HI) studies have been particularly powerful in capturing these events \citep{Rots1990, Yun1994, Hulst2005, Morganti2006, Oosterloo2010, Serra}. Further evidence for gas accretion is given by deep \HI\ observations of the gaseous halos of star-forming galaxies \citep{Oosterloo2007, Fraternali2002, Heald2011}. These studies have detected significant amounts (for example, 30$\%$ of the total \HI\ mass in NGC 891, \citealt{Oosterloo2007}) of gas at large vertical distances from the disk. However, most of the halo gas is thought to be of galactic fountain origin, with a few features (counter-rotating clouds, etc.) that point to a contribution from external accretion.

When looking for signatures of accretion, peculiar \HI\ gas distribution and kinematics such as scattered clouds, tails, bridges, ring-like structures, warped disks, lopsidedness may point to ongoing gas accumulation events. However, in many cases these signatures only provide indirect evidence for accretion. 
An alternative approach for testing the growth history of galaxies is provided by multiwavelength analysis. Merger events can produce stellar remnants around galaxies, e.g., streams, shells, like the ones observed in M 31\citep{Ibata2001} and other nearby systems \citep{Hernquist1988, Malin1997, Duc2015}. Even though such remnants do not directly indicate gas accumulation, deep optical observations of low surface brightness stellar components can be highly complementary for the \HI\ data. 
The drawback is that these \HI\ and optical studies of individual galaxies are very expensive in terms of observing time because sensitive, high resolution observations are needed to detect associated faint features, especially at larger distances. Furthermore, unsettled \HI\ signatures will disappear in a few Gyr \citep{vanWoerden1983, Serra2006}, making it difficult to draw a conclusion on the origin of the gas. 
In the upcoming era, the next generation of radio telescopes will provide resolved \HI\ data for hundreds of thousands of galaxies thanks to the Square Kilometre Array (SKA) pathfinders and precursors \citep{Oosterloo2010b, DeBoer2009, Booth2009, Fernandez2015}. These surveys will provide measurements for galaxies at $z > 0.2$, currently unexplored with \HI\ observations, yielding unparalleled samples for testing gas accretion processes and extending them to cosmologically significant distances. This requires a method to select, statistically, the best candidates for having recently accreted their gas reservoirs using global \HI\ scaling relations.

Early \HI\ studies have explored variations in the \HI\ content of galaxies as function of morphological type, size, and environment (e.g. \citealt{Haynes1984, Knapp1985, Roberts1994}). Correlations between \HI\ content and observed galaxy properties do exist, as cold gas provides the building block for stars. If large gas reservoirs are linked to galaxies with low SF activity, indicating unusually inefficient star-formation, this might suggest that the system has recently accreted fresh supplies of gas. To be able to identify such cases, first we need to establish `\HI\ normalcy' for galaxies given their SF/structural properties. In recent years, blind, single-dish surveys like the \HI\ Parkes All Sky Survey (HIPASS, \citealt{Meyer2004, Zwaan2005}), and the Arecibo Legacy Fast ALFA survey (ALFALFA, \citealt{Giovanelli2005, Haynes2011}) have provided \HI\ observations for tens of thousands of galaxies that can be used to study the global \HI\ scaling relations \citep{Huang2012, Denes2014} in the nearby Universe ($z < 0.06$ for ALFALFA). However, these studies are biased toward the most gas-rich detections due to the surveys being shallow. The GALEX Arecibo SDSS Survey (GASS; Catinella et al. 2010, 2013) was specifically designed to perform deep (gas-fraction limited), targeted \HI\ observations for an unbiased sample of $\sim$800 galaxies, selected by stellar mass ($>$ 10$^{10}$ \msun) and redshift (0.025 $<$ z $<$ 0.05). These multiwavelength observations made it possible to measure the correlation between \HI\ content and global quantities such as structural/SF properties. \cite{Catinella} show that the linear combination of NUV$-r$ color and stellar mass surface density ($\mu_{\star}$) is one of the best predictors of the \HI\ gas fraction (the ratio of \HI\ to stellar mass, $M_{\rm HI} / M_{\star}$). The correlation is not surprising, since the NUV$-r$ color and $\mu_{\star}$ are established proxies for specific star formation rate (SFR/M$_{\star}$) and morphological type respectively, properties shown to be closely tied to gas fraction (see also \citealt{Fabello2011, Gereb2015, Brown2015}). Galaxies with unusually high gas content for their SF properties would be displaced above the relation, showing an excess amount of \HI\ compared to the average galaxy population. Once the candidates are identified, detailed studies of the resolved \HI\ and stellar/SF properties are needed to gain a detailed understanding of the accumulation of the excess gas.%

Using the GASS sample we selected one of the most intriguing cases of an \HI-excess galaxy, GASS 3505. In this paper we carry out a multiwavelength analysis on GASS 3505 with the goal of gaining insights into the formation of the galaxy. Optical Sloan Digital Sky Survey (SDSS; \citealt{York}) and ultraviolet (UV) Galaxy Evolution Explorer (GALEX; \citealt{Martin2005}) observations are available for this galaxy by selection. These data allow for measurements of the stellar and SF properties of GASS 3505. We also obtained interferometric \HI\ observations with the Very Large Array (VLA) to look at the resolved \HI\ distribution, kinematics, and physical conditions of the gas, e.g. \HI\ column density. Combining these resolved \HI\ observations with the UV data allows us to study GASS 3505 in terms of the global SF laws \citep{Schmidt1959, Kennicutt1989, Kennicutt1998}. Deep optical imaging was carried out at the Apache Point Observatory (APO), allowing us to trace faint stellar components that would be missed by the shallow SDSS observations. We search for signatures of accretion within the galaxy using the resolved \HI\ data and deep optical imaging, and we study the availability of \HI\ for accretion by exploring the local (Mpc) environment of GASS 3505. Finally, we combine the observational results with smoothed particle hydrodynamic (SPH) and N-body simulations to provide a scenario for the formation of this exceptional object. The best model is chosen based on its ability to reproduce the global properties of GASS 3505. 

In this paper the standard cosmological model is used, with parameters $\Omega_{\rm M}$ = 0.3, $\Omega_{\Lambda}$ = 0.7 and $H_0$ = 70 km s$^{-1}$ Mpc$^{-1}$.

%% file: Tex_files/GASS3505_Selection.tex
\begin{figure}
\begin{center}
\includegraphics[width=.3\textwidth]{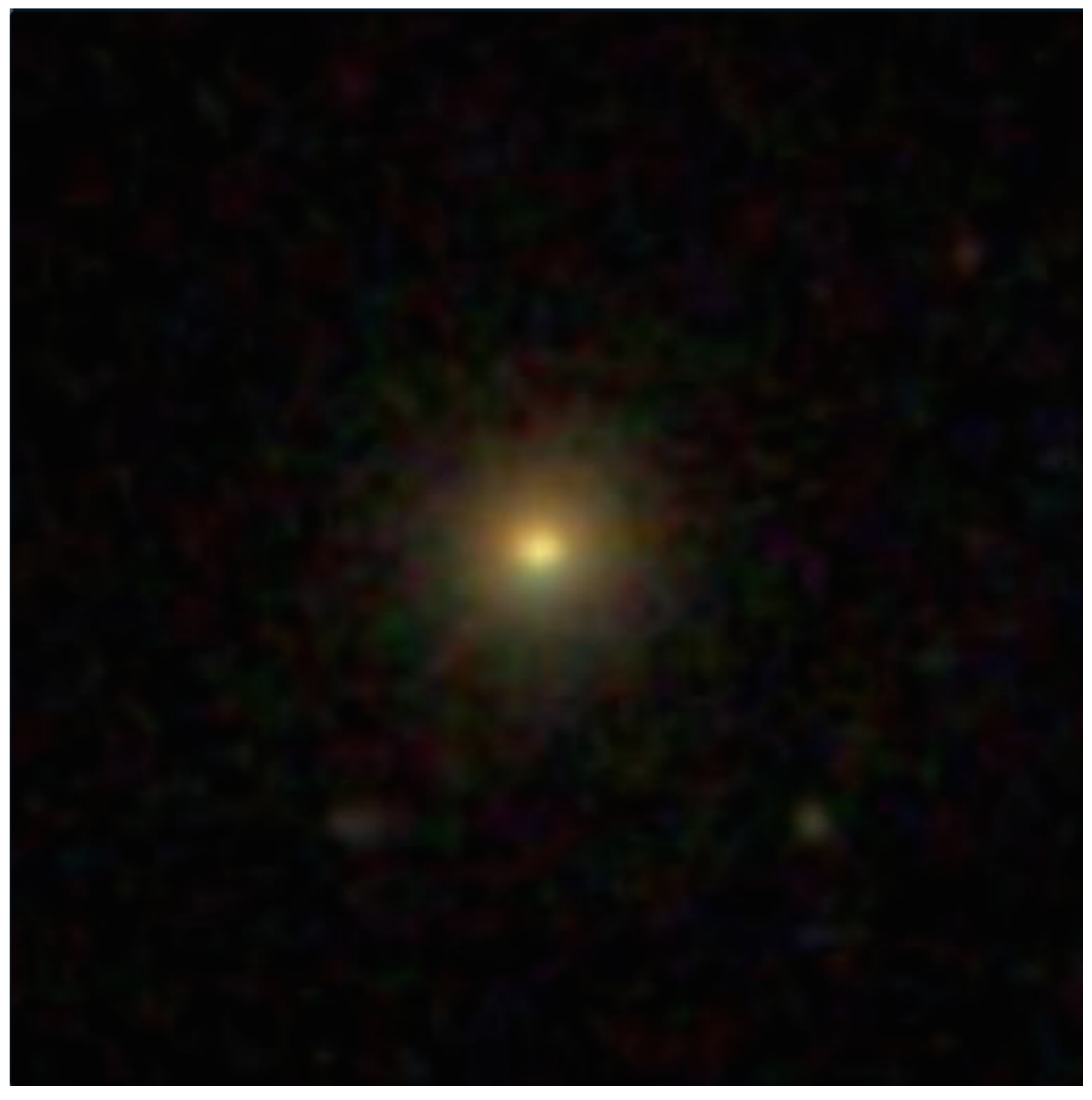}
\caption{SDSS optical image of GASS 3505 (50 arcsec diameter).}\label{fig:SDSSimage}
\end{center}
\end{figure}

\section{The selection of GASS 3505} \label{Sec:Selection}

GASS 3505 is selected in the framework of the multiwavelenth GASS survey of massive ($M_{\star} > 10^{10}$ \msun), nearby galaxies \citep{Catinella}. Molecular gas observations of GASS 3505 have been carried out by the COLD GASS survey \citep{Saintonge2011}, producing an upper limit for the molecular gas fraction. The multiwavelength properties of the galaxy are summerized in \mbox{Table \ref{table:GASS3505}}. The stellar mass is from the Max Planck Institute for Astrophysics (MPA)/Johns Hopkins University (JHU) value-added catalogs, based on SDSS DR7. The measurement assumes a \cite{Chabrier2003} initial mass function, and it is accurate to better than 30 percent. The stellar mass surface density is defined as $\mu_{\star}$ = $M_\star$ /(2$\pi$ $r_{50,z}^2$), where $r_{50,z}$ is the radius containing 50$\%$ of the Petrosian flux in the SDSS $z$ band. Its color and high stellar mass surface density ($\mu_{\star}$) suggests that GASS 3505 is a red sequence, early-type galaxy, or in other words a quiescent object in terms of its SF properties. This is also supported by the SDSS optical image (Fig. \ref{fig:SDSSimage}) and spectrum (not shown) of the galaxy. Given its structural and SF properties, it is somewhat surprising that the GASS survey observations detected $10^{9.9}$ \msun\ of \HI\ mass in GASS 3505. 

\begin{figure}
\begin{center}
\includegraphics[width=.4\textwidth]{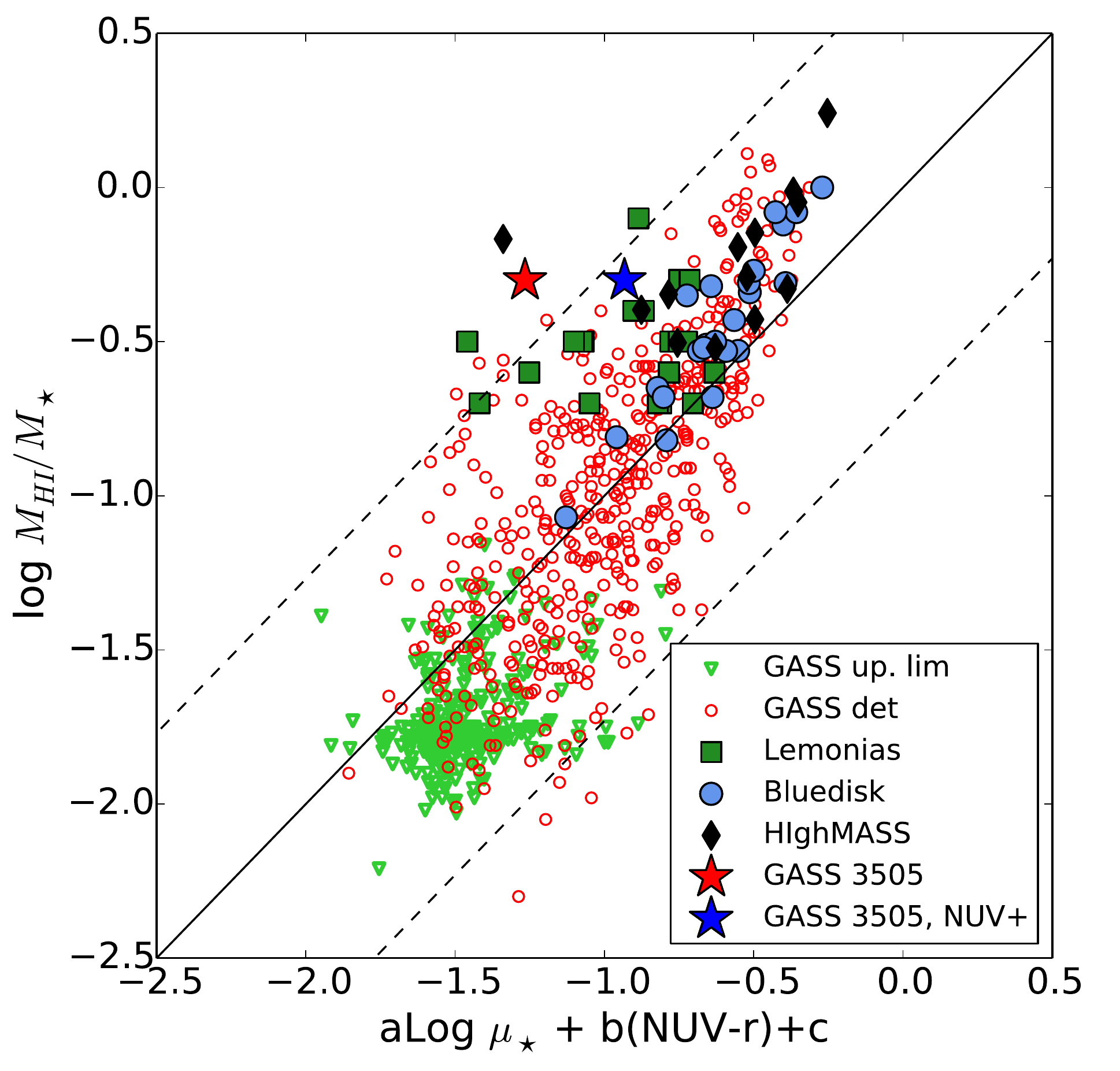}
\caption{\HI\ gas fraction plane. Red circles and green triangles represent GASS detections and non-detections, respectively. GASS 3505 is marked by a red star. The blue star indicates GASS 3505 after we remeasure its location as described in Sec. \ref{sec:UVdata}. The solid line represents the 1:1 relation, and the dashed lines the 2.5$\sigma$ deviation from it. The values of the coefficients are a = $-0.240$, b = $-0.250$, c = $2.083$. Other surveys indicated in the legend are discussed in Sec. \ref{Sec:Context}.}\label{fig:gf_plane}
\end{center}
\end{figure}

\begin{figure}
\begin{center}
\includegraphics[width=.4\textwidth]{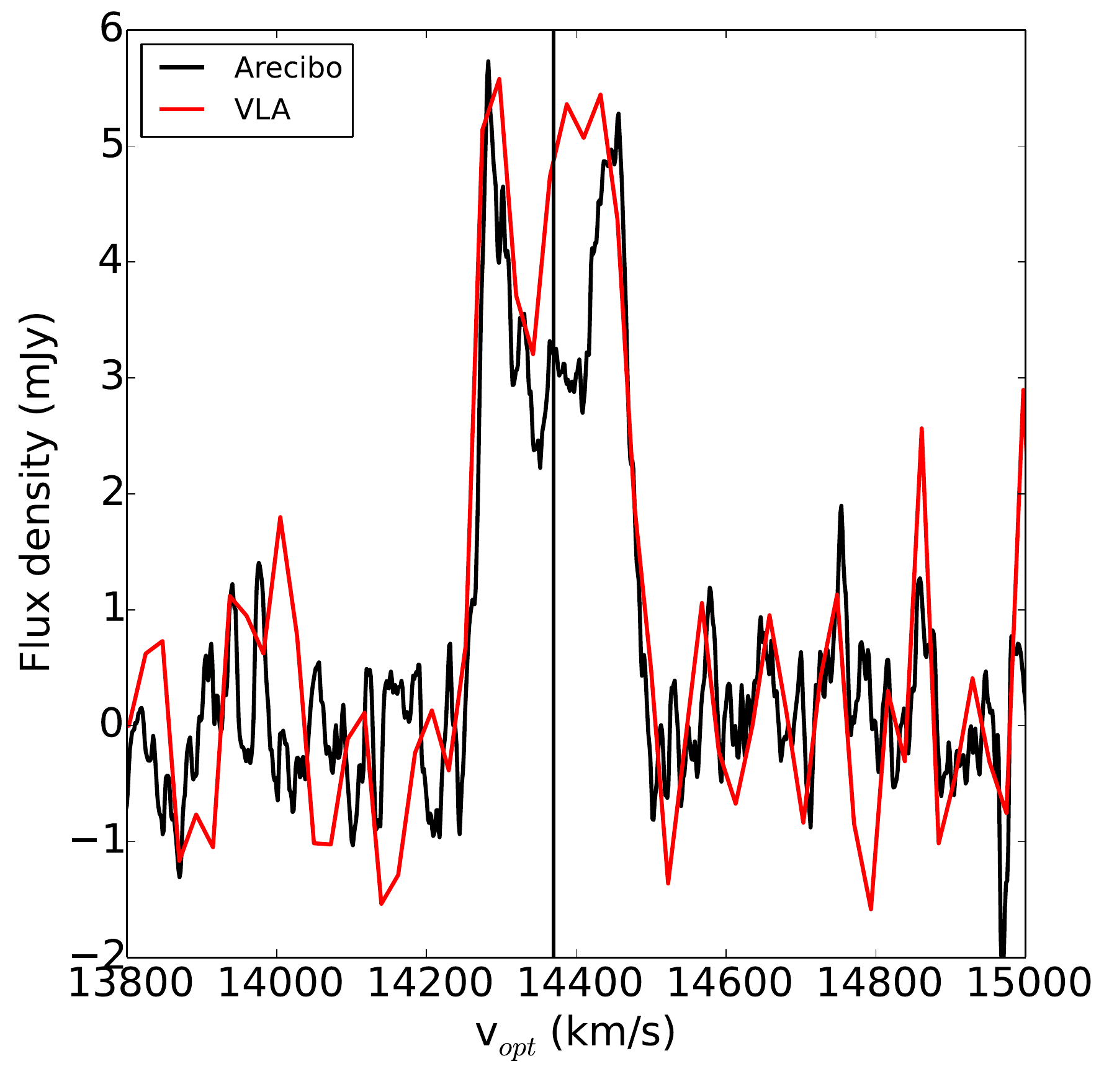}
\caption{Arecibo spectrum (black), overlaid with the VLA \HI\ spectrum of GASS 3505 (red). The black vertical line indicates the systemic velocity corresponding to the SDSS redshift.}\label{fig:GASS3505_HIline}
\end{center}
\end{figure}

In Fig. \ref{fig:gf_plane} we show the best fit relation between the measured \HI\ mass fraction and that expected from the NUV$ - r$ color/stellar mass surface density for the GASS sample, which we refer to as the {\it gas fraction plane} \citep{Catinella2013}. The deviation from the plane on the y-axis was shown to be a good proxy for the \HI-deficiency/excess parameter of galaxies \citep{Cortese2011}. GASS 3505 is marked by a red star in \mbox{Fig. \ref{fig:gf_plane}}. It was first identified as the strongest outlier within the GASS sample, with a striking $\delta = 3.3\sigma$ deviation (the scatter of the plane is $\sim$0.3 dex) from the average relation. The gas fraction plane constitutes a relation between the \HI\ content of galaxies and their star-forming/morphological properties, therefore the high deviation of GASS 3505 from the plane represents a contrast between its \HI-richness and lack of star-formation. One can note that there are only a handful of objects with similar gas excess in \mbox{Fig. \ref{fig:gf_plane}}. It can be shown that \HI-excess galaxies with $>$2.5$\sigma$ deviation (a conservative limit for selecting outliers, see dashed lines in Fig. \ref{fig:gf_plane} for the GASS sample) on the y axis are an extremely rare population, with a volume density of $\sim$2 $\times 10^{-5}$ Mpc$^{-3}$. This number is based on galaxies from both GASS and ALFALFA (meeting the $M_{\star} > 10^{10}$ \msun\ GASS selection criteria) that are not confused within the Arecibo beam; the volume refers to the region enclosed by the ALFALFA and GASS surveys in the redshift range $0.01 < z < 0.05$. The low volume density suggests that GASS 3505 is a unique object in terms of its \HI\ and SF properties.

\begin{figure*}
\begin{center}
\includegraphics[width=1.\textwidth]{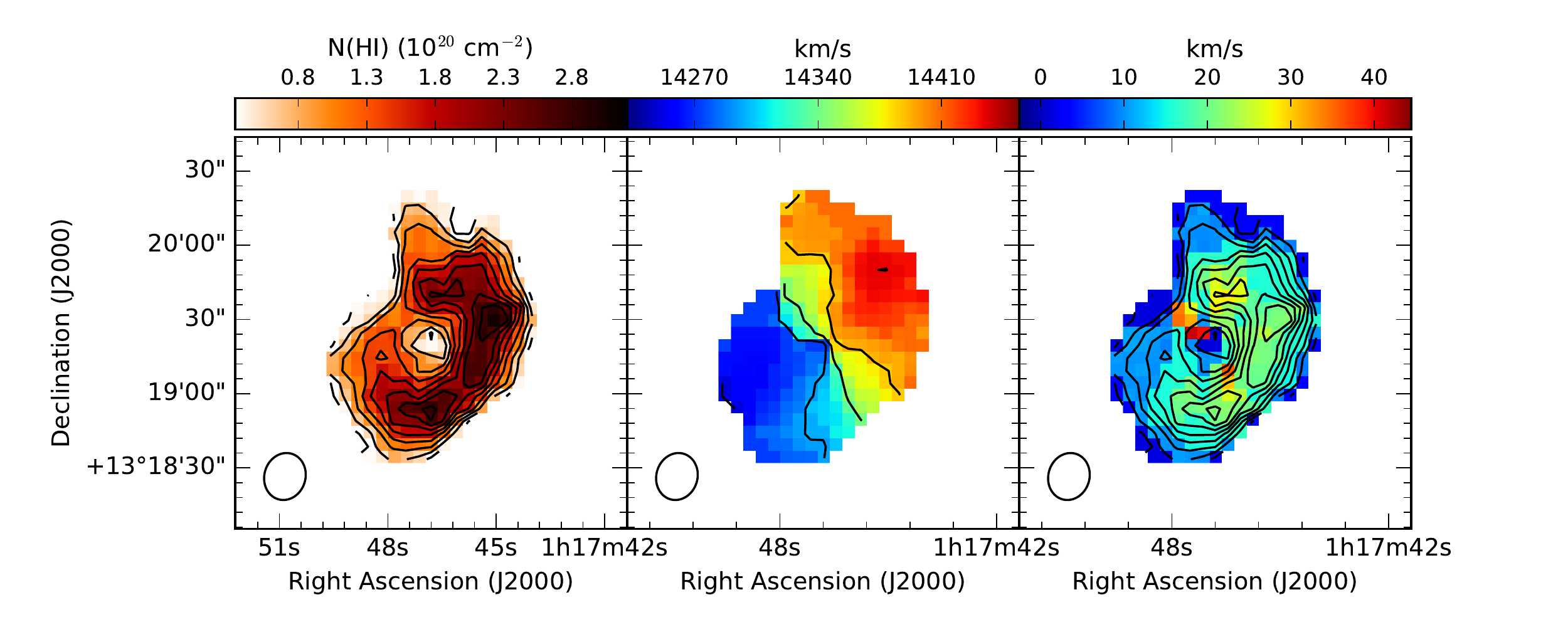}
\caption{VLA \HI\ column density map (left panel), first-moment (velocity field, middle panel) and second-moment (velocity dispersion, right panel) maps of GASS 3505. The \HI\ column density contours range between 3.4 $\times$ $10^{19}$ cm$^{-2}$ (3$\sigma$) and peak intensity (3 $\times$ $10^{20}$ cm$^{-2}$), with steps of 3$\sigma$. The beam size is shown in the bottom left corner of each panel.}\label{fig:GASS3505}
\end{center}
\end{figure*}

\begin{table}
  \caption{Basic properties of GASS 3505. Row.1 SDSS ID; Row.2 SDSS spectroscopic redshift; Row.3 base-10 logarithm of the stellar mass surface density; Row.4 NUV $-$ $r$ color, corrected for Galactic extinction; Row.5 base-10 logarithm of the \HI\ mass; Row.6 base-10 logarithm of the stellar mass; Row.7 \HI\ gas fraction measured by GASS; Row.8 deviation from the gas fraction plane ($\delta$); Row.9 base-10 logarithm of the molecular gas mass upper limit; Row.10 molecular gas fraction upper limit. }\label{table:GASS3505}
  
   \begin{center}
     \begin{tabular}{l c c}		
    &  & References\\ 	
\hline									     
SDSS ID & J011746.76+131924.5 & 1\\ 
z   & 0.047932 & 1\\
log($\mu _{\star}$) [M$_{\odot}$ kpc$^{-2}$] & 8.83 & 1 \\
NUV $-$ $r$ [mag] & 4.92 & 1 \\
log($M_{\rm \HI}$)  [\msun] & 9.91 & 1 \\
log($M_{\star}$) [\msun] & 10.21 & 1 \\
$M_{\rm \HI}$/$M_{\star}$ & 0.50 & 1 \\
$\delta$ & 3.3 & 1 \\
log($M_{\rm H2}$) [M$_{\odot}$] & $<$ 8.93  & 2\\
$M_{\rm H2}$/$M_{\star}$  & $<$ 0.05 & 2 \\

\hline
\end{tabular}
     {\bf References.} (1) Catinella et al. (2010); (2) Saintonge et al. (2011)
\end{center}
\end{table}

In order to understand such outliers from the gas fraction plane, we need to study these systems in detail and determine how they acquired their gas - was it recently accreted from the environment, or has the gas stayed in equilibrium for Gyrs without forming stars? Due to the large, 3.5 arcmin beam of the Arecibo single dish telescope, the GASS survey does not provide spatially resolved information on the \HI\ gas distribution and kinematic properties. Thus, interferometric observations are needed to study the morphology and kinematics of the gas, which we can then relate to different gas accumulation processes.

%% file: Tex_files/HI_obs.tex
\section{Interferometric HI observations}\label{HI_observations}

The \HI\ observations of GASS 3505 were carried out using the combination of the Very Large Array (VLA) and Karl G. Jansky VLA (JVLA) antennas during the transition phase of the telescope. In August-September 2009, the target was observed for a total of 24 hours in C configuration. The 6 MHz bandwidth is covered by 62 frequency channels, with a corresponding channel width of \mbox{$\sim$20 km s$^{-1}$}. The data were calibrated and reduced using the Common Astronomy Software Applications (CASA, \citealt{McMullin2007}) package, following standard procedures. The data were cleaned of RFI, and the continuum was subtracted by fitting a low order polynomial to the line-free channels. The \HI\ data cube was created with natural weighting to maximise the signal-to-noise ratio using the task {\sl clean}, and it was Hanning smoothed. After Hanning smoothing, the 3$\sigma$ rms noise is $\sim$0.2 mJy beam$^{-1}$, and the velocity resolution is $\sim$40 km s$^{-1}$. The synthesised beam of the data cube is 19.2" $\times$ 16.6".

For other \HI\ sources in the cube, which were detected at a different redshift than GASS 3505, the continuum was subtracted by excluding the corresponding \HI\ lines. For these sources we created separate, primary-beam-corrected \HI\ cubes to account for falling sensitivity at larger radii from the central galaxy. The \HI\ moment maps presented in this paper were created by selecting the channels with \HI\ emission for each source. In the selected channels the \HI\ emission was masked at the 3$\sigma$ level. The images were smoothed to 20" ($\sim$19 kpc) to enhance faint, diffuse emission in the \HI\ distribution. The final moment maps, i.e. integrated intensity (zero-moment), velocity field (first-moment), and velocity dispersion (second-moment) were created from the masked data cubes.

In Fig. \ref{fig:GASS3505_HIline} we show the \HI\ spectrum of GASS 3505. Small differences between the VLA and Arecibo spectra arise due to the different velocity and spatial resolution of the two datasets. We derive the \HI\ mass to be log$(M_{\rm \HI})$  = $9.88 \pm 0.05$ \msun\ based on the integrated intensity map, confirming the results of the GASS observations. In Fig \ref{fig:GASS3505} (left panel), the gas reveals a ring morphology in the column density map (or disk with gas column densities below our N(\HI) $<$ 3.4 $\times$ $10^{19}$ cm$^{-2}$ 3$\sigma$ detection limit in the central region), extending $\sim$50 kpc beyond the galaxy center. The distribution of the gas is uneven, showing an \HI\ depression in the eastern part of the ring. The highest column density peaks of N(\HI) $=(3 \pm 0.02) \times 10^{20}$ cm$^{-2}$ are measured in the southern and western part. These are the regions where the conditions for star formation would be more favorable. The effect of the gas on the SF properties will be discussed in Sec. \ref{HI_SF}.

The velocity field of the ring displays regular kinematics in Fig. \ref{fig:GASS3505} (middle panel). The second-moment map (right panel) shows that the velocity dispersion is normal for a disk \citep{Pickering1997}. However, a direct measurement of the velocity dispersion is limited by the coarse velocity resolution of our data. 
We derive the maximum rotational velocity within a 50 arcsec radius to be $153 \pm 12$ km s$^{-1}$ using the inclination $i$ = 42.8$^{^{\circ}}$ estimated based on the minor-to-major axis ratio of the \HI\ ring, and a position angle PA = 317$^{\circ}$ derived based on the major axis (measured counter clockwise from north through east, where the north direction corresponds to $0^{\circ}$). 

These results show that the large amount of gas in GASS 3505 is settled into an extended, rotating configuration. The regular \HI\ kinematics suggests that the gas must have completed at least 1-2 orbital periods \citep{vanWoerden1983, Serra2006}. For 50 kpc radius and $153$ \kms\ rotational velocity, the \mbox{T = $(2 \pi  $r)/v} orbital period is $\sim2$ Gyr, suggesting that the ring can not be younger than this age.

%% file: Tex_files/Optical_data.tex
\begin{figure*}
\begin{center}
\includegraphics[width=0.99\textwidth]{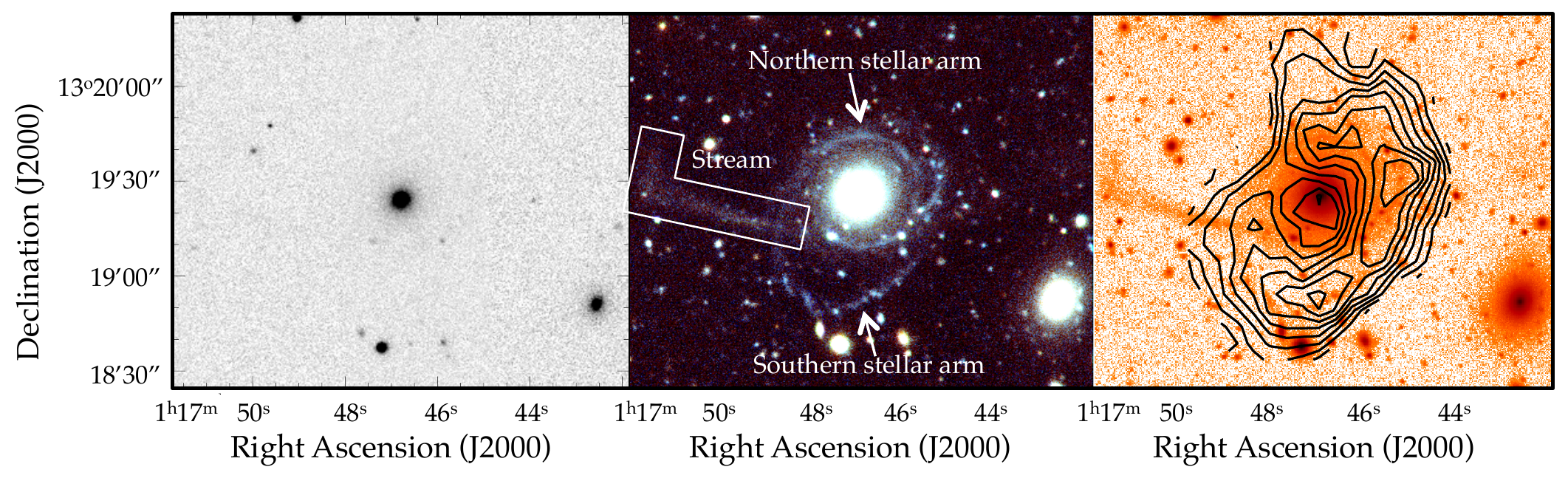}
\caption{{\it Left panel}: SDSS $r$-band image of GASS 3505. {\it Middle panel}: Deep APO color image of GASS 3505. The faint stellar stream and arms are indicated in white. {\it Right panel:} Deep $r$-band APO image of GASS 3505 overlaid with the VLA \HI\ column density contours. The \HI\ contours range between 3.4 $\times$ $10^{19}$ cm$^{-2}$ (3$\sigma$) and peak intensity (3 $\times$ $10^{20}$ cm$^{-2}$) with steps of 3$\sigma$, as in Fig. \ref{fig:GASS3505} }\label{fig:Gass3505_color}
\end{center}
\end{figure*}

\section{Optical data: Evidence for stellar accretion} \label{Sec:deepimaging}

Here we make use of the $g$ and $r$ band deep optical imaging to look for faint stellar features in GASS 3505. The galaxy was observed with the SPICAM imager on the 3.5m telescope at Apache Point, on 4 October, 2010. SPICAM has a field of view of 4.8 arcmin, and we employed 2$\times$2 binning for all exposures resulting in a pixel size of 0.282\arcsec. The galaxy was observed for a total of 2hrs in {\it r} and 1.5hrs in {\it g} band, under seeing conditions of 0.85\arcsec ({\it r}-band) and 1\arcsec ({\it g}-band). The observations were divided into individual exposures of 10 min each. We used a five-point dither pattern with exposures taken centered on the galaxy and at $\pm$30 arcsecond spacings in RA and Dec. Individual exposures were flat-fielded using dome flats, and then aligned and median-combined using Swarp \citep{Bertin2002}. The final combined images have been flux calibrated by comparing the magnitudes of 5 bright (non saturated) stars from our images with the values obtained from SDSS images. The typical uncertainty in the flux calibration is of the order of $\sim$0.03 mag.

\begin{figure}
\begin{center}
\includegraphics[width=.4\textwidth]{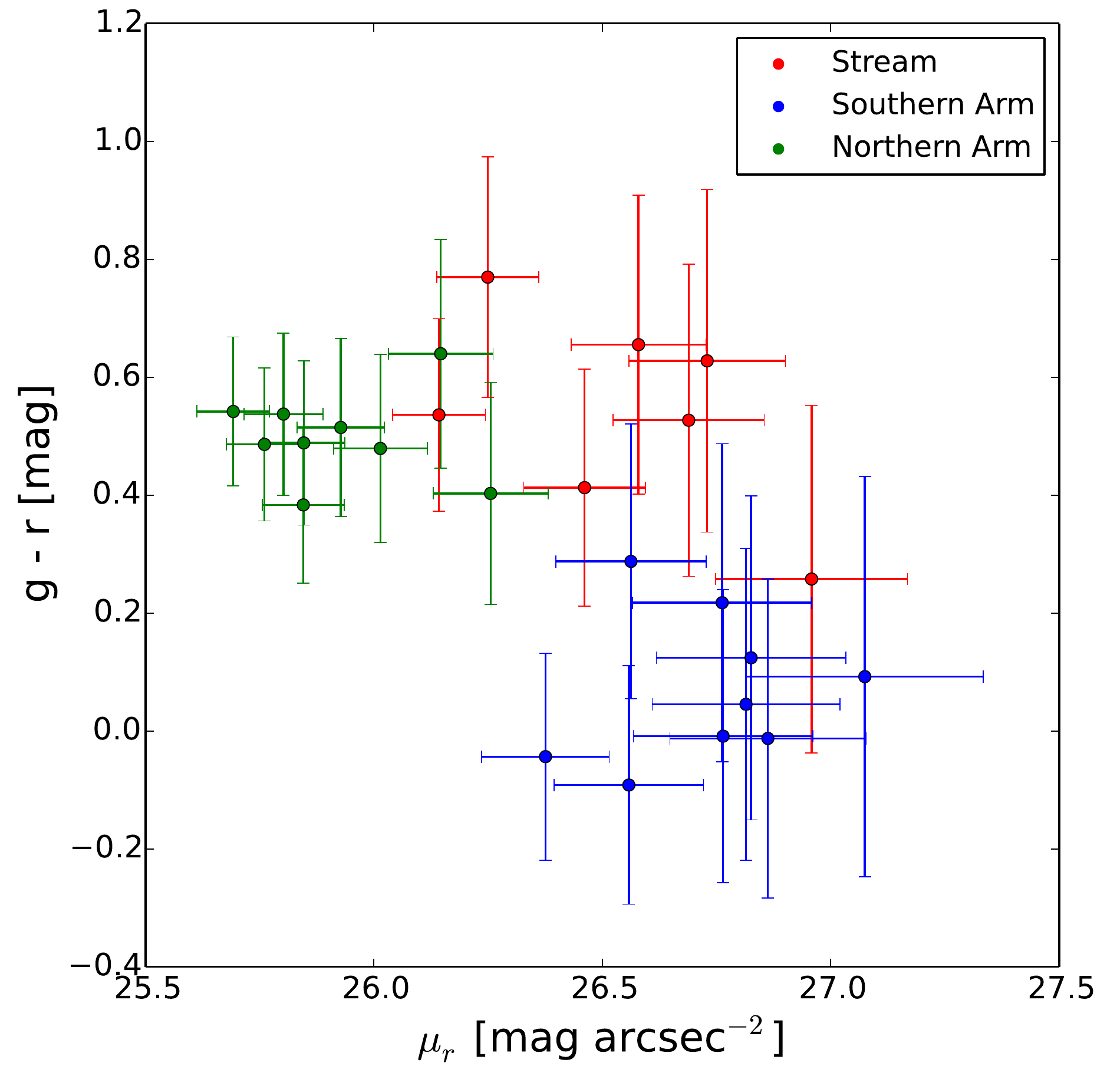}
\caption{APO $g-r$ color vs. $r$-band surface brightness measured in the stream, southern, and northern arm of GASS 3505, as indicated in the legend.}\label{fig:SB_color}
\end{center}
\end{figure}

In Fig. \ref{fig:Gass3505_color} we show the SDSS $r$-band (left panel) and APO color image (central panel). In the shallower SDSS image GASS 3505 appears as an early-type galaxy. Interestingly, deep imaging reveals extended, low surface-brightness stellar emission around the central bulge. We detect two stellar arms embedded in the \HI\ ring, and a stellar stream that extends $\sim$60 kpc beyond the galaxy centre. We measure the photometry of the low surface brightness regions in 2 arcsec radius apertures, centred on the southern, the northern stellar arm, and on the stream. The $g-r$ color and $r$-band surface brightness distribution is shown in Fig. \ref{fig:SB_color}. 
The two stellar arms are bluer than the stream in general, and they coincide with the highest \HI\ column density regions in Fig \ref{fig:Gass3505_color} (right panel). This may suggest that the formation of the stellar arms is related to the presence of gas. The analysis of the star-forming properties in these regions is presented in Sec. \ref{HI_SF}.

Stellar streams are the typical signatures of galaxy remnants disrupted in merger events like in our local neighbour, M31 \citep{Ibata2001, McConnachie2003}. Such tidal features are thought to be a manifestation of galaxy growth in the present day Universe. We create a polygonal aperture with a surface area of 613 ${\rm arcsec}^{2}$ (depicted in Fig. \ref{fig:Gass3505_color}, middle panel) in order to measure the stellar mass content of the stream in GASS 3505. We estimate the stellar mass to be $(2.7 \pm 0.9) \times $10$^{8}$ M$_{\odot}$, following the method of \cite{Zibetti2009}, accounting for $\sim$2$\%$ of the bulge stellar mass. In Fig. \ref{fig:SB_color} we show that part of the stellar stream (3 out of 7 data points) is almost as red as the central bulge ($g - r$ = 0.82\footnote{The APO color measurement is consistent with the SDSS photometry} mag). In addition, we show in Fig. \ref{fig:Gass3505_color} (right panel) that no \HI\ gas is detected in the stellar stream. We argue that the stellar stream in GASS 3505 is the mark of recent interaction like in M31, confirming that the galaxy has experienced some kind of merger event. Possible scenarios for the formation of the stream in relation to the gas content will be discussed in Sec. \ref{Sec:Discussion_sim}.

%% file: Tex_files/UV_data.tex
\section{UV data: The SF properties of GASS 3505} \label{sec:UVdata}

\begin{figure*}
\begin{center}
\includegraphics[width=.8\textwidth]{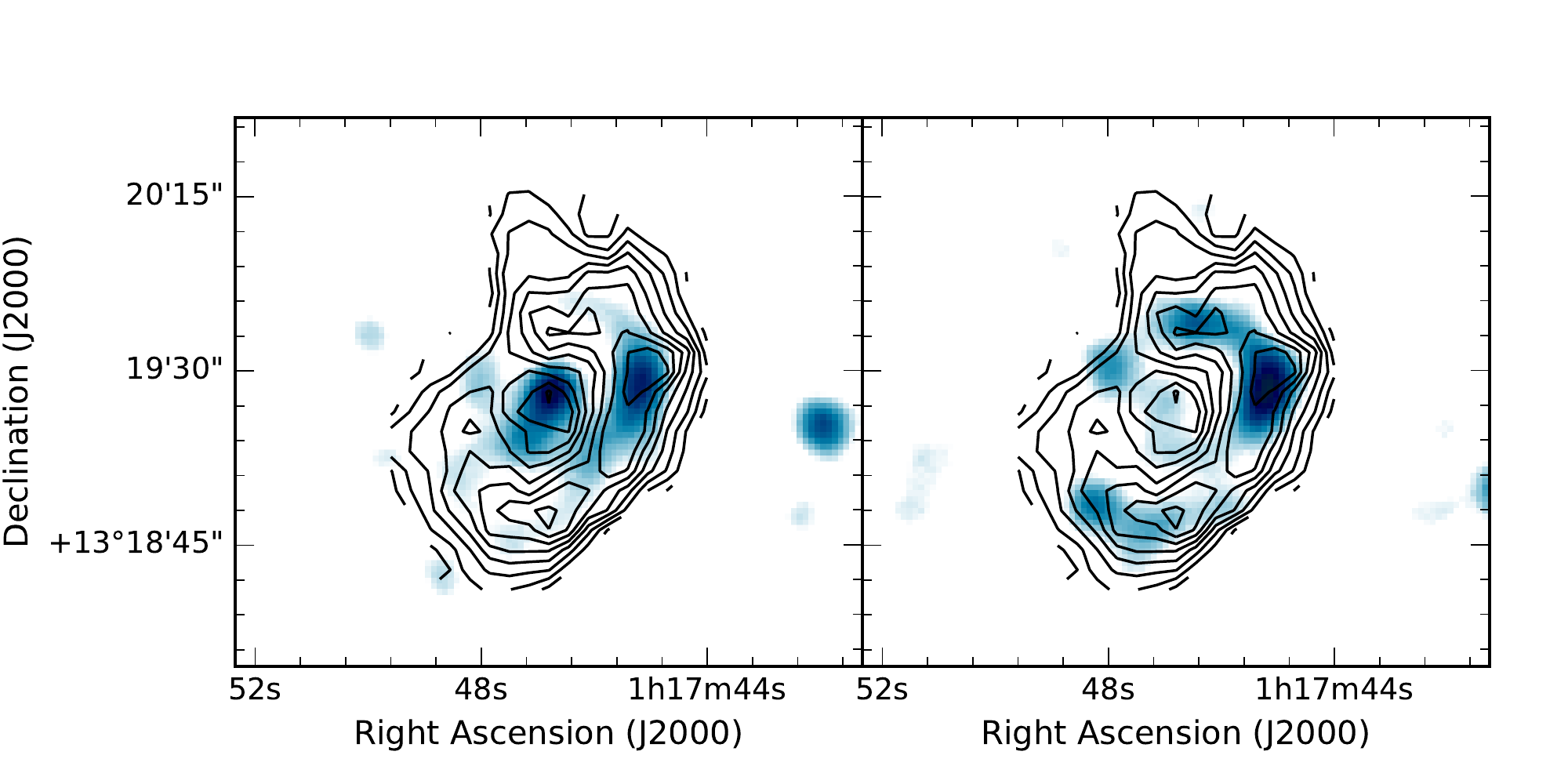}
\caption{{\it Left panel}: GALEX NUV image of GASS 3505 smoothed to 13 arcsec resolution (only for illustration purposes), overlaid with the VLA N(\HI) contours (black) as in Fig. \ref{fig:GASS3505}. {\it Right panel}: GALEX FUV smoothed (13 arcsec) image, overlaid with the same contours.}\label{fig:Galex}
\end{center}
\end{figure*}

We use the GALEX Multimission Archive at Space Telescope Science Institute (MAST) to retrieve FUV (1350-1750 ${\AA}$) and NUV (1750-2800 ${\AA}$) images for GASS 3505. The resolution of the extracted NUV and FUV maps are 5.6" and 4" respectively. The maps are background subtracted using IRAF, and the photon counts are converted into flux and magnitudes as described on the GALEX webpage\footnote{$http://galexgi.gsfc.nasa.gov/docs/galex/FAQ/counts\_ background.html$}.

In Fig. \ref{fig:Galex} (left panel), the NUV map reveals emission both from the old stellar population of the central bulge, and from the SF regions in the disk. The color measurement using NUV and $r$ band magnitudes reported in \cite{Catinella} was restricted to the optical radius measured from SDSS, hence it only provides a lower limit on the total emission. The missing light can contribute to the large deviation of GASS 3505 from the gas fraction plane, and here we want to correct for this effect. We use the APO $r$ band and GALEX NUV map to remeasure the location of GASS 3505 on the scaling relations, accounting for the emission of the faint extended disk. We integrate the NUV and $r$ band emission within an aperture of 50", and re-calculate the color of GASS 3505 after correcting for Galactic extinction. 
Following \cite{Wyder2007}, the extinction correction is ${\rm A}_{NUV} -{\rm A}_r = 1.98{\rm A}_r$, where ${\rm A}_r$ is the $r$ band extinction from SDSS. We estimate the color to be ${\rm NUV} - r = 3.59$, where the error on the color measurement is $\sim$0.05 mag. The result is shown in Fig. \ref{fig:gf_plane}, where the new measurement of GASS 3505 is marked by a blue star. The deviation of this point from the plane is 2.15$\sigma$. Thus, even if we take into account the stellar counterpart of the faint disk, the deviation of GASS 3505 is still significant.

We use the FUV band as the best tracer of current star formation available to us. In Fig \ref{fig:Galex} (right panel), the GALEX FUV image shows that ongoing SF is present at the outskirts of the galaxy in a ring-like structure, indicating the presence of young stellar populations. This is consistent with the optical measurements in Sec. \ref{Sec:deepimaging}, where we show that the color of the low surface brightness region is bluer than the bulge. The FUV flux is converted into SFR following \cite{Bigiel2010}. We expect the dust-to-gas ratio to be low at the outskirts of the galaxy, therefore corrections for internal extinction are neglected. We derive a total star formation rate of SFR$_{\rm FUV}$ = $0.13 \pm 0.01$ M$_{\odot}$ yr$^{-1}$ inside a 50" radius. As expected, the level of SF is orders of magnitudes lower than that of normal SF galaxies \citep{Kennicutt1989,  Kennicutt1998}. Next we study the effect of low \HI\ surface density on the further SF/evolutionary processes of the galaxy using the resolved FUV and \HI\ maps.

%% file: Tex_files/K_S.tex
\subsection{Inefficient SF in low HI surface density galaxies}\label{HI_SF}

\cite{Bigiel2010} show that the \HI-dominated outskirts (1$-$2$\times$r$_{25}$) of galaxies populate a new parameter space at the low end of the $\Sigma_{\rm SFR}$ vs. $\Sigma_{\rm gas}$ relation (also known as the Kennicutt-Schmidt relation; \citealt{Schmidt1959, Kennicutt1989, Kennicutt1998}). In this regime the efficiency of SF is very low, with gas depletion timescales longer than a Hubble time ($>$ 10$^{10}$ yr). \cite{Bigiel2010} argue that the column density of \HI\ is perhaps the key factor in regulating the SFR in outer galaxy disks. In GASS 3505, \HI\ represents the dominant gas mass component at large radii (see Table \ref{table:GASS3505}), hence the outskirts of the galaxy are clearly \HI-dominated (Fig. \ref{fig:Gass3505_color}), similarly to the \cite{Bigiel2010} sample. In the following analysis we examine the SF fuelling processes in GASS 3505, focusing on the connection between atomic gas content and FUV emission.

\begin{figure}
\begin{center}
\includegraphics[width=.4\textwidth]{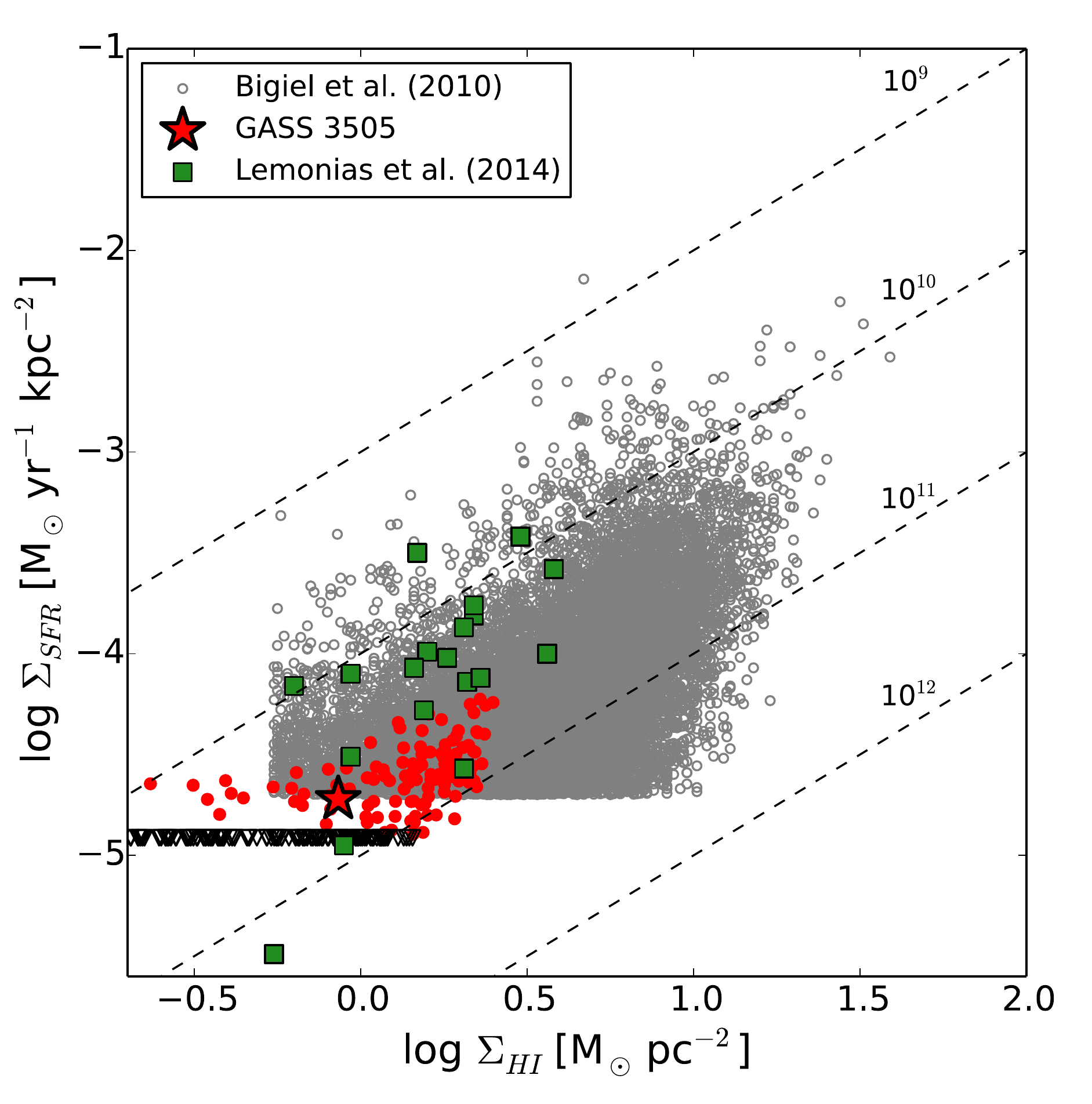}
\caption{Pixel-by-pixel analysis of the SFR surface density vs. \HI\ surface density in GASS 3505. Star-forming regions (where the FUV emission is above 3$\sigma$) are marked by red dots, and non-star-forming regions (3$\sigma$ FUV upper limit) are shown as black triangles. We also show the Bigiel et al. (2010) and Lemonias et al. (2014) samples in grey and green points (respectively) as a comparison. Constant gas depletion timescales of 10$^{9}$ $-$ 10$^{12}$ yr are marked by black dashed lines. }\label{fig:K_S}
\end{center}
\end{figure}

In Fig. \ref{fig:Galex} (right panel), the GALEX FUV map is overlaid with the \HI\ column density contours. The highest \HI\ column density ($>$ $10^{20}$ cm$^{-2}$) regions of the \HI\ ring show a clear spatial correlation with FUV emission from star forming regions. In order to learn about the connection between \HI\ and SFR surface density, we carry out a pixel-by-pixel analysis in the star-forming and quiescent regions of GASS 3505. For this analysis we smooth the FUV image to 20" to match the spatial resolution of the \HI\ map. The \HI\ column density is converted into surface density according to 1 M$_{\odot}$ pc$^{-2}$ = 1.3 $\times$ 10$^{20}$ atoms cm$^{-2}$ \citep{Hulst1987}. In Fig. \ref{fig:K_S}, star-forming regions are represented by pixels with FUV emission above 3$\sigma$ significance. The \HI\ surface density is measured at the corresponding pixels in the \HI\ map, i.e., within the 3$\sigma$ FUV mask. These points are marked by red dots in Fig. \ref{fig:K_S}, whereas non-SF regions are marked by black triangles. The latter points correspond to the 3$\sigma$ FUV upper limit outside of the SF regions, measured in pixels where \HI\ is still detected above the 3$\sigma$ level. We also show the global measurement, marked by a red star, where both the \HI\ and FUV emission are integrated within a circular aperture of 50". For a comparison, grey circles indicate the \cite{Bigiel2010} results for the outskirts of spirals and dwarf galaxies. The cutoff in the \cite{Bigiel2010} distribution is due to a 3$\sigma$ detection limit in their published data. It should be noted that the pixel size in our \HI\ map is 5 arcsec, corresponding to $\sim$4.7 kpc at the redshift of the galaxy, which is larger than the $\sim1$ kpc resolution of the \cite{Bigiel2010} observations.
The black dashed lines in Fig. \ref{fig:K_S} correspond to constant gas depletion timescales of 10$^{9}$ $-$ 10$^{12}$ years. The long gas consumption timescale of $\sim$10$^{11}$ yr in GASS 3505 suggests that the gas disk can provide fuel for low-level SF over more than a Hubble time. This is likely due to the physical conditions of the gas, implying that the efficiency of SF is regulated by \HI\ gas density at the low end of the SF law, as previously proposed by \cite{Bigiel2010}. 

At the current SF efficiency the disk will likely remain \HI-dominated, unless environmental effects become important in removing/stripping the gas. Studying the environment is also important for our understanding of the gas accumulation processes in GASS 3505. To investigate whether gas-rich satellite accretion would be a viable process for building up the gas reservoir in GASS 3505, as a next step we examine the Mpc-scale environment around the galaxy.

%% file: Tex_files/Environment.tex
\section{The environment of GASS 3505}\label{Sec:Environement}
 
\begin{figure*}
\includegraphics[angle=90,width=.64\textwidth]{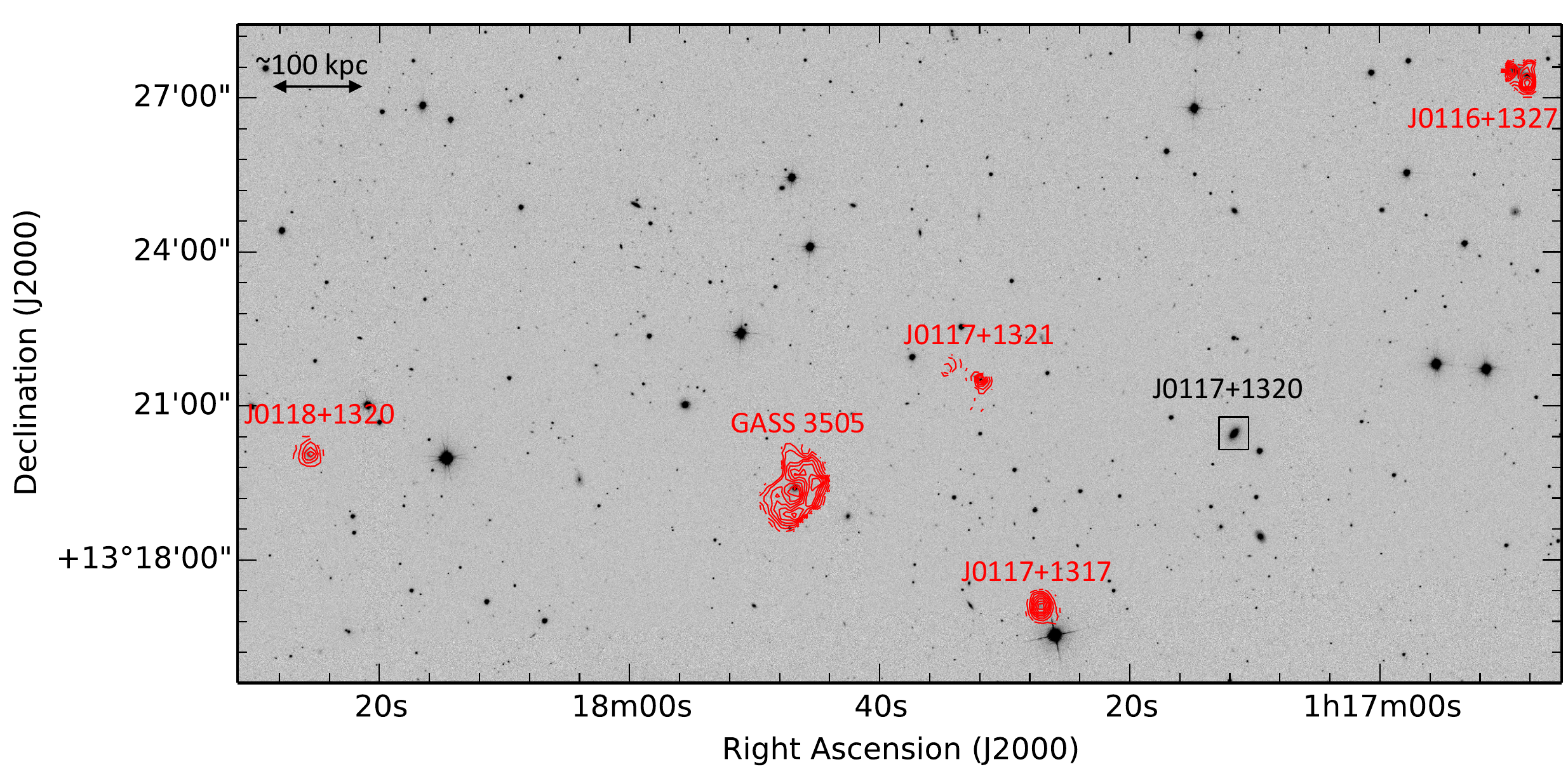}
\begin{center}\caption{SDSS $r$-band optical image (grey), overlayed with the VLA \HI\ column density contours (red) of the detections. In the observed VLA volume one additional SDSS galaxy is found with an \HI\ non-detection, marked by a black square.}\label{fig:Otpical+NHI}
\end{center}
\end{figure*}

Using the \HI\ and SDSS data we study the Mpc-scale environment surrounding GASS 3505 and explore the availability of gas in nearby galaxies. We identify four \HI\ detections in the vicinity of GASS 3505, and we summarize their properties in Table \ref{table:HIparameters}. The Mpc environment of the VLA \HI\ field is shown in Fig. \ref{fig:Otpical+NHI}, overlaid on the SDSS $r$-band optical image. The mentioned sources are identified based on their \HI\ detections, and we look for further galaxies (\HI\ non-detections) in the Mpc environment using SDSS spectroscopic information. We find one early-type galaxy at 0.51 Mpc projected distance from GASS 3505 at $z = 0.048484$, by SDSS identifier J011711.65+132027.3 (marked as J0117+1320 in Fig. \ref{fig:Otpical+NHI}). The stellar mass of this galaxy is log$(M_\star)$ = 10.80 \msun, and it is not detected in \HI\ at the upper limit of \mbox{\mhi\ $<$ $10^{8}$ M$_{\odot}$} (assuming a line width of 200 km s$^{-1}$). 

We explore the possibility that these surrounding galaxies are in gravitational interaction with GASS 3505. Based on the Virial theorem, a galaxy is considered gravitationally bound if the potential energy of the system is larger than its total kinetic energy, i.e., if $(G \times M)/r_{\rm sep} > 1/2 \times v^{2}$, where $G$ is the gravitational constant, $M$ is the total mass of GASS 3505 and its potential satellite, $r_{\rm sep}$ is the physical separation between the pair, and $v$ is the velocity difference between these two. The three-dimensional velocity field can be approximated as $v \simeq \sqrt{3} v_{\rm los}$ if we assume that the system is isotropic, where $v_{\rm los}$ is the line-of-sight velocity difference, the only measurable velocity component available to us \citep{Evslin2014}. The $r_{\rm sep}$ and $v_{\rm los}$ measurements are presented in Table \ref{table:HIparameters}. The total mass is calculated as the sum of $M_{\rm halo} + M_{\star} + M_{\rm HI}$, where $M_{\rm halo}$ is estimated based on the stellar mass of each object following the study by \cite{Behroozi2013}. We find that J0118+1320 is gravitationally bound based on this criterion, whereas the other galaxies are not part of the system. 
There is no clear evidence of interaction detected between GASS 3505 and its companion, J0118+1320. On the other hand, other galaxies are interacting with each other, i.e., we detect a gas-rich merger, and an edge-on spiral galaxy with a warped \HI\ disk close to its companion.  These results suggest a picture where GASS 3505 is the central galaxy of a pair system, located in a low density, but dynamic environment where interactions between galaxies are taking place. Such a low-density, gas-rich environment provides favorable conditions for the build-up/maintenance of the extended gas ring, which would not be possible in denser environments. In the next section we investigate how this exceptional object relates to other \HI-rich samples and early-type galaxies.

%% file: Tex_files/Context_otherSurveys.tex
\section{GASS 3505 in the context of other surveys}\label{Sec:Context}

There exist a number of other surveys from recent years focused on \HI-rich\ galaxies. In this Section we discuss GASS 3505 in the context of these studies.

\subsection{The HI-rich samples}

The Bluedisk study by \cite{Wang2013} focuses on understanding the gas properties in galaxies predicted to be \HI-rich for their stellar mass and color. Resolved \HI\ observations were obtained for 23 selected targets, which have been used in combination with optical and UV data to study the morphology of the gas with respect to star-forming properties at the outskirts of these galaxies. The Bluedisk sample mainly consists of late-type galaxies with large \HI-to-optical size ratios (compared to control samples), where young star-forming regions are populating the outer disks.

The HighMass sample \citep{Huang2014} includes the most \HI-rich galaxies detected by the blind ALFALFA survey. Briefly, these were selected to be the most \HI-rich objects for their stellar masses in the $\alpha$.40 sample \citep{Haynes2011}, with $M_{\rm HI}$ > $10^{10}$ \msun, and with a gas fraction ($M_{\rm HI}$/$M_\star$) that is more than 1$\sigma$ above the running average in a given $M_\star$ bin (see Fig. 1 of \citealt{Huang2014}). In total, 34 galaxies from $\alpha$.40 meet the HIghMass selection criteria. GASS 3505 is just below the HIghMass selection cut off with its $10^{9.9}$ M$_{\odot}$ of \HI\ mass and 50$\%$ gas fraction. \cite{Huang2014} show that HIghMass galaxies exhibit healthy ongoing star formation and, at least some of them, are `normal' star-forming galaxies scaled-up to higher \HI\ mass and gas fraction \citep{Hallenbeck2014}.

In a study by \cite{Lemonias2014}, the GASS and ALFALFA samples are used to select \HI-rich galaxies with gas fractions in the top $5\%$ of the GASS distribution. These galaxies have high gas fractions for their stellar masses, and are \HI-massive with $M_{\rm HI}$ > $10^{10}$ \msun. Within this sample, \cite{Lemonias2014} identify a population where high \HI\ masses are associated with unexpectedly low specific SFR-s. They show that low surface density gas is likely the reason behind the inefficient SF in these galaxies.

In the following analysis we compare GASS 3505 to the mentioned \HI-rich samples in terms of the global scaling relations. The Bluedisk and \cite{Lemonias2014} sample are massive galaxies with $M_\star$ $>$ $10^{10}$ \msun\ like GASS 3505, and for consistency we only consider HIghMass galaxies which meet this high stellar mass selection criteria. The NUV $- r$ color information for Bluedisk and the \cite{Lemonias2014} galaxies are taken from their published tables. For the HIghMass sample we extract the NUV magnitudes from GALEX and correct the NUV $-$ $r$ color for Galactic extinction following \cite{Wyder2007}. In total 13 HIghMass galaxies have available NUV information and are more massive than $M_{\star}$ $>$ 10$^{10}$ \msun. 

In Fig. \ref{fig:gf_plane} we show these \HI-rich populations on the gas fraction plane along with the GASS sample. Unlike the \HI-excess galaxy GASS 3505, which is a clear outlier, most of the \HI-rich galaxies follow the global scaling relations. Given that the majority of Bluedisk and HIghMass samples were shown to be late-type galaxies with `healthy' star-forming properties, it is not surprising that these galaxies lie on our gas fraction plane. On the other hand, among the \cite{Lemonias2014} sample many galaxies have deviations of $\sim2 \sigma$. Because low gas density is suggested to be responsible for regulating the SF processes in these galaxies, we return to the Kennicutt-Schmidt law to investigate the reason behind the location of GASS 3505 on the gas fraction plane compared to the \cite{Lemonias2014} sample. The difference is clearly illustrated in Fig. \ref{fig:K_S}, where GASS 3505 is located in a lower \HI\ and SFR surface density regime compared to the general population of the \cite{Lemonias2014} galaxies. These results support the idea that the high deviation of GASS 3505 from the plane is the result of the galaxy having too much \HI\ relative to what is expected given that it is in a quiescent phase. Thus, the main difference between the \HI-excess GASS 3505 and other \HI-rich galaxies manifests in the distinct SF properties of the galaxies compared to their \HI\ mass fractions. This is due purely to sample selection.

\subsection{\HI-rich early-type galaxies}

It has been known for several decades that many early-type galaxies do contain \HI\ gas \citep{Knapp1985, Wardle1985, Driel1991, Gorkom1997, Sadler2002, Morganti2006, Oosterloo2010, Serra}. Among observations, \atlas\ is a representative early-type galaxy sample with high-resolution, sensitive \HI\ data. 

The \atlas\ survey has found that about 50$\%$ of the \HI\ detections in early-type galaxies are distributed in a regularly rotating disk/ring morphology. The largest \HI\ structures extend up to many tens of kpc-s with \HI\ masses between $10^{8}-10^{10}$ \msun. The \HI\ column density in the majority of the \atlas\ detections is below 10$^{20}$ cm$^{-2}$, with the highest column density being N(\HI) = 3$\times$10$^{20}$ cm$^{-2}$. Our results in Sec. \ref{HI_observations} show that GASS 3505 is tracing the highest end of the early-type \HI\ mass and column density distribution. 
However, the greatest difference stands in the measured gas fraction of the two samples. The \HI\ gas fraction of GASS 3505 is at least 10 times greater than the average gas fraction in \atlas\ galaxies (Serra, private communication). Only one galaxy in \atlas\ has a comparable \HI-to-stellar mass ratio, however the \HI\ in this object is unsettled. This suggests that GASS 3505 and the \atlas\ sample have likely very little in common. We conclude that GASS 3505 is not just a `normal' early-type galaxy with \HI, but it is a rare object with exceptionally high gas content.

%% file: Tex_files/Simulation.tex
\begin{figure*}
\begin{center}
\includegraphics[width=.6\textwidth]{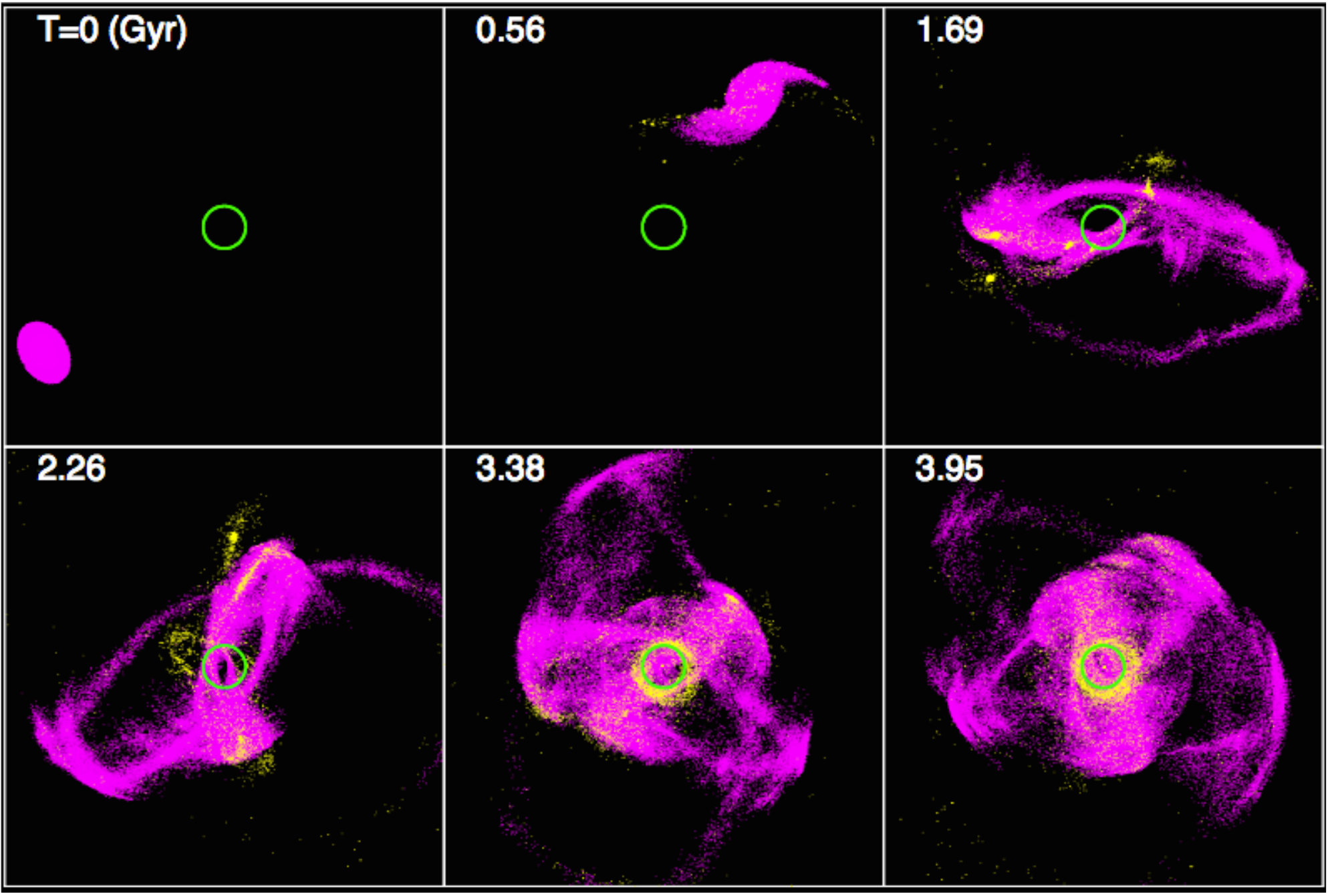}
\caption{Snapshots of the merger between the ellitpical and disk galaxy. Magenta colors indicate old stars, new stars are shown in yellow. The central green circle represents the $r_{\rm 90}$ (6 kpc) of the early-type bulge. Time steps are indicated on the top of the images in Gyr.}\label{fig:Simulations_stars}
\end{center}
\end{figure*}

\begin{figure*}
\begin{center}
\includegraphics[width=.6\textwidth]{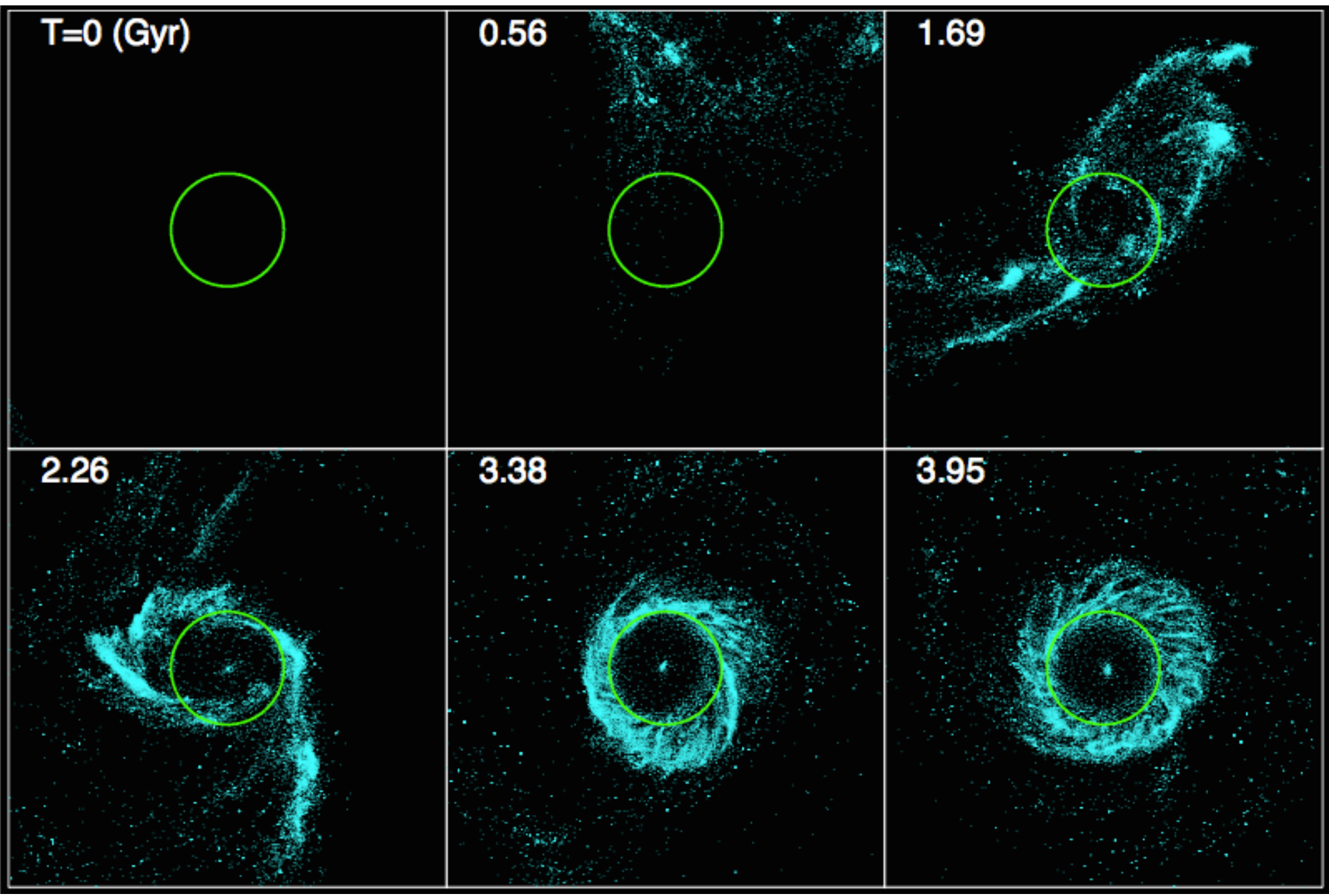}
\caption{XY snapshot for gas only, zoomed-in version}\label{fig:Simulations_gas} 
\end{center}
\end{figure*}  
\section{The formation scenario of GASS 3505} \label{Sec:Discussion_sim}

The observational results show that GASS 3505 is a rare galaxy in terms of its gas fraction and relative SF properties. The presence of the stellar stream is a strong evidence that the bulge of GASS 3505 has accreted a companion. In this Section we discuss possible scenarios for the origin of the huge \HI\ reservoir in GASS 3505 in an attempt to answer the following question: has all of the \HI\ been brought in by the progenitor of the stellar stream, or other gas accumulation scenarios have to be involved?

\subsection{Numerical model: gas-rich satellite accretion?} 
 
We perform a chemodynamical simulation of a merger between the bulge of GASS 3505 and a gas-rich disk galaxy. We use the chemodynamical code by \cite{Bekki2013} to simulate the time evolution of the SFR and gas content in a merger between an elliptical and a spiral galaxy (`E-Sp' merger). 
The dark matter halo and stellar bulge of the elliptical galaxy are similar to that of GASS 3505, however the simulated elliptical galaxy has no gas. The accreted companion is composed of a dark matter halo, a stellar disk, and a gaseous disk with no bulge component, like the Large Magellanic Cloud (LMC). The remnant appears (almost) completely destroyed in the observations, suggesting that the progenitor had to be a low stellar-mass ($\sim 10^{9}$ \msun) object compared to GASS 3505. We assume that the disk galaxy has 10$^{10}$ \msun\ of \HI, equivalent to the measured gas content. We do not incorporate molecular hydrogen (${\rm H_2}$) calculations into the present simulation. This is because no molecular gas is detected in GASS 3505 at the $M_{\rm H_2}$/$M_{\star} <$ 0.05 upper limit, and our main interest is in studying the physical properties of the \HI\ and stellar components. More details on the modelling of the gas and stars can be found in \cite{Bekki2013}.

\begin{table}
\centering
\begin{minipage}{85mm}
\caption{
A list of model parameters for the fiducial model in the chemodynamical simulation. Row.1 total dark matter mass of the elliptical galaxy; Row.2 total stellar mass of the elliptical galaxy; Row.3 initial size of the stellar spheroid in the elliptical galaxy; Row.4 total dark matter mass of the disk galaxy; Row.5 total stellar matter mass of the disk galaxy; Row.6 total gas mass of the disk galaxy; Row.7 initial stellar disk size of the disk galaxy; Row.8 orbital eccentricity of the E-Sp merging; Row.9 initial separation of the E-Sp interaction; Row.10 pericenter distance of the E-Sp merging; Row.11 initial inclination angle between the stellar disk of the disk galaxy and the orbital plane of the disk galaxy in the E-Sp merging; Row.12 threshold gas density for star formation; Row.13 initial gaseous metallicity of the disk galaxy; Row.14 mass resolution of the disk galaxy; Row.15 spatial resolution of the disk galaxy.}\label{table:Sim}
\begin{tabular}{l c}
{Model parameters } & {Adopted values}  \\
\hline
$M_{\rm E, dm}$ [${\rm M}_{\odot}$] & $10^{12} $  \\

$M_{\rm E, s}$ [${\rm M}_{\odot}$] & $6 \times 10^{10} $  \\

$R_{\rm E, s}$ [kpc] & $8.6$   \\

$M_{\rm D, dm}$ [${\rm M}_{\odot}$]& $1.7 \times 10^{11} $  \\

$M_{\rm D, s}$ [${\rm M}_{\odot}$] & $1.0 \times 10^{9} $  \\

$M_{\rm D, g}$ [${\rm M}_{\odot}$] & $1.0 \times 10^{10} $  \\

$R_{\rm D, s}$ [kpc] & $10.5$  \\

$e_{\rm p}$ & 0.7  \\

$R_{\rm i}$ [kpc]&  70   \\

$R_{\rm p}$ [kpc] &  17.5   \\

$\theta$ [deg] &  45  \\

$\rho_{\rm th}$ [atom cm$^{-3}$] &  1  \\

${\rm [Fe/H]_0}$ [dex] &  $-0.38$  \\

$m_{\rm s}$  [${\rm M}_{\odot}$] &  $10^4$  \\

$\epsilon_{\rm s}$ [pc] &  $152$   \\
\end{tabular}
\end{minipage}
\end{table}

When using the observed values for setting the initial conditions, the accreted system is destroyed in the inner region of the central galaxy and the simulation produces a compact \HI\ disk instead of a ring. 
We ran $20+$ simulations to find the best model that can explain the observed properties of GASS 3505. In Table \ref{table:Sim} we show the results of the `fiducial' model. The main difference compared to observations is that in our best model, the $M_{\rm E, s} = 6 \times 10^{10}$ \msun\ stellar mass of the bulge seems significantly higher than the $1.6 \times 10^{10}$ \msun\ measured stellar mass (the error on the measurement is 30$\%$). We note that the measured stellar mass would increase to $2.8 \times 10^{10}$ \msun\ if we assumed a Salpeter initial mass funtion (IMF, \citealt{Salpeter1955, Gallazzi2008}) instead of the current \cite{Chabrier2003}. The uncertainty applies only to the observations because in the simulation we use a fixed stellar mass without the need for assumptions regarding the IMF. Furthermore, the low surface brightness features revealed by our deep imaging suggest that our stellar mass (which is based on SDSS data) might be underestimated. Unfortunately, it is impossible for us to quantify such a bias, as disentangling between features belonging to GASS 3505 and foreground/background objects is extremely challenging. Thus, the discrepancy in observed vs. simulated stellar mass ratios can either be the result of the mentioned uncertainties, or other gas accretion mechanisms have to be involved in the formation of such massive, extended disks. Nevertheless, we show the results of the model because it can reproduce the general properties of the galaxy, e.g., gas distribution and kinematics, SFR, stellar morphology. 

In Fig. \ref{fig:Simulations_stars} we show the evolution of the accreted stellar component during the merger. It takes about $\sim$1 Gyr for the first streams to form, and by the end of the simulation the accreted galaxy is completely destroyed. The time evolution of the gas is presented in Fig. \ref{fig:Simulations_gas}. At the early stages the gas is scattered around the galaxy, and it settles into a ring in about $\sim$2 Gyr. New stars are forming slowly in the disk (yellow points in Fig. \ref{fig:Simulations_stars}), with a maximum SFR of 0.4 M$_{\odot}$yr$^{-1}$. 

In the literature one can find other examples for early-type galaxies which are thought to have formed via gas-rich mergers \citep{Veron1995, Morganti1997, Balcells2001, Serra2006}. So what makes GASS 3505 special?
Our simulation suggests that very specific mass ratios/orbital parameters are needed in order to produce an \HI-excess ring. The accreted galaxy in the simulation is a LMC-type galaxy with a $M_{\rm HI}$/$M_{\star}$ = 10 gas fraction. This galaxy is an outlier on the global \HI\ scaling relations, meaning that it is a rare object \citep{Huang2012}. Interactions between such \HI-rich, low-mass objects and early-type galaxies are even more rare, as the two types preferentially reside in different environments \citep{Dressler1980, Lewis2002, Gomez2003, Blanton2009}. The combination of these effects would explain the low number density of GASS 3505-type galaxies in the nearby Universe. On the other hand, the simulated scenario is unable to reproduce the observed gas properties of GASS 3505 if the mass discrepancy between the best model and observed galaxy properties is considered meaningful. This would be an indication that additional gas accumulation scenarios have to be involved in the build-up of the \HI\ ring. 

\subsection{Other sources of gas}

 \noindent
{\sl Multiple, gas-rich galaxy accretion:}\\
Our simulation shows that very specific orbital parameters/progenitors are needed in order for the gas to be "trapped" at the outskirts of the galaxy, and to build extended \HI\ disks that are similar to GASS 3505. We can not fully exclude the possibility of multiple mergers in the past, although our results suggest that it is very unlikely that a second interaction would not trigger SF in the gas.\\

\noindent
{\sl Cold or hot mode accretion:} \\
Another possibility is that the \HI\ ring could have formed via cold or hot mode accretion. However, this scenario can not explain why GASS 3505-type \HI-excess galaxies are so rare in the nearby Universe, as these processes should be more common in galaxies. Given the mass discrepancy in our simulation we can not rule out that some part of the \HI\ was accreted via cold or hot mode accretion, however it is unlikely that these processes alone are responsible for building such exceptionally gas-rich disks.

%% file: Tex_files/Discussion.tex
\section{Summary and Conclusions}

The analysis presented in this paper has allowed us to develop a better understanding of the origins of the \HI-excess in GASS 3505. Resolved VLA observations show that the large \HI\ reservoir (10$^{9.9}$ \msun) is distributed in a 50 kpc gas ring in this galaxy. UV and deep optical imaging reveal a low surface brightness, younger stellar population embedded in the \HI\ gas, extending tens of kpcs beyond the early-type body of GASS 3505. The SF associated with the \HI\ is very inefficient, with a global SFR of \mbox{$\sim$0.1 M$_{\odot}$ yr$^{-1}$}. Throughout the entire disk a good spatial correlation is observed between the \HI\ and SF surface densities. This, along with the fact that no molecular gas is detected in GASS 3505, suggests that the \HI\ column density is likely the main regulating factor for the SF processes. We estimate that it would take more than a Hubble time to consume the large \HI\ reservoir by SF processes at the current level of efficiency.

The star-forming \HI\ ring in GASS 3505 is connected to a more complex stellar stream, which is a typical signature of stellar accretion events. As was previously described by \cite{Hernquist1988, Johnston2001, Duc2015}, faint stellar features are important reminders that the morphology and evolution of early-type galaxies could be more complex than previously thought, and that deep imaging is essential for identifying related accretion events. GASS 3505 is a typical case where, based on the presence of the stellar stream, one might expect to have an unsettled \HI\ distribution. Instead, the gas morphology is significantly more regular than that of the stars. This example shows that multiwavelength analysis is crucial for the success of gas accretion studies.  

We provide a theoretical basis for the formation scenario of GASS 3505 using SPH and N-body simulation by combining observations of the structural and kinematic properties of the gas, the stellar properties, and information on the local environment. We find that it is possible to form a GASS 3505-type galaxy via a merger between a bulge and a gas-rich low-mass galaxy. However, we find a discrepancy between the modeled and observed stellar masses, in the sense that the bulge mass in the best model is larger than the observed value. This discrepancy can either be the result of the large uncertainties on the compared stellar masses (i.e., IMF assumed in the observations, mass content of the outer disk), or it can suggest that other accretion mechanisms have to be involved in the formation of the disk, i.e., cold/hot more accretion or a series of gas-rich merger events.

Gas accretion is likely responsible for re-activating star formation in galaxies that otherwise would be passively evolving, quiescent objects. This is a channel that would confirm the idea that some galaxies may be moving slightly away from the red sequence, e.g. passive ellipticals that are currently building up an \HI-rich disk \citep{Cortese2009}. Given that 40$\%$ of early-type galaxies in the field contain \HI\ \citep{Serra}, the mentioned gas accretion events could play a major role in the build-up of the \HI\ reservoirs and further evolution of early-type and transitional objects.

Future surveys that will be carried out with the next generation of radio telescopes like the Australian Square Kilometre Array Pathfinder (ASKAP, \citealt{DeBoer2009, Duffy2012}), Apertif \citep{Oosterloo2010b}, MeerKat \citep{Booth2009}, Karl G. Jansky Very Large Array (JVLA, e.g. CHILES survey, \citealt{Fernandez2015}) will extend \HI\ studies to cosmologically significant distances. These will likely provide measurements of the most massive and gas-rich systems in the distant Universe, objects which will be excellent probes for studies of gas accretion in galaxies. The HIGHz study at $z \sim 0.2$ supports the assertion that some of these higher redshift galaxies are indeed very \HI-rich\ \citep{Catinella2008, Catinella2015}. Thus, it is important to prepare for the necessary work at higher redshift using local studies. The first step is to increase the local \HI-excess galaxy sample size. In future work, we aim to do a statistical analysis of the general population of \HI-excess galaxies using resolved \HI, deep optical, UV, and integral field unit (IFU) observations combined with numerical modelling.

%% file: Tex_files/Appendix.tex
\section{\HI\ environment} \label{Sec:app}

\begin{table*}  
   \begin{center}

   \caption{SDSS and \HI\ parameters of the detected sources in the environment of GASS 3505. Col.1 SDSS identifier; Col.2 SDSS-based short ID used in this paper. Col.3 redshift; Col4. projected distance of galaxy from GASS 3505; Col.5 physical distance of galaxy from GASS 3505 calculated based on the luminosity distance and angular separation; Col.6 relative line-of-sight velocity difference between galaxy and GASS 3505; Col.7 Color; Col.8 Stellar mass. Where indicated, the values are taken from the SDSS catalogue, otherwise we derive the stellar mass based on the color and \HI\ redshift. Col.9 \HI\ mass; Col.10 \HI\ column density peak; Col.11 SFR extracted from SDSS; Col.12 Specific SFR (SFR M$_{\star}^{-1}$) extracted from SDSS;}\label{table:HIparameters}

   \begin{threeparttable}

 \scalebox{0.9}{  

     \begin{tabular}{l c c c c c c c c c c c}
\hline															     

 SDSS ID &  ID &  $z$               & $d_{\rm proj.}$ & $r_{\rm sep}$  & $v_{\rm los}$ & $g - r$   & log($M_{\star}$) & log($M_{\rm HI}$) &  N(\HI)$_{peak}$ & log(SFR)  &  log(SSFR) \\ 
         &  in this paper &             &Mpc   & Mpc & \kms\ & mag   & M$_{\odot}$      & $M_{\odot}$  & 10$^{20}$ cm$^{-2}$ & M$_{\odot}$ yr$^{-1}$ & yr$^{-1}$\\ 
\hline
J011825.66+132003.7 & J0118+1320 &0.04804$^{\star}$   & 0.56 & 0.75 & 32.4 &0.58 & 9.42                & 9.2  &3.4 &  -      & - \\
J011731.96+132130.9 & J0117+1321 &0.04868             & 0.25 & 3.45  & 224.2 &0.66 & 9.79$^{\dagger}$    & 9.3  &4.8 &  -0.30 & -10.14  \\ 
J011727.16+131706.2 & J0117+1317 &0.04941$^{\star}$   & 0.32 & 6.80 & 443.1 &0.17 & 8.35                & 9.5  &6.2 &  -      & - \\ 
J011649.16+132730.5 & J0116+1327 &0.04881             & 0.95 & 4.15  & 263.2 &0.41 & 9.71$^{\dagger}$     & 9.7  &6.4 &  0.11 & -9.65\\ 

\end{tabular}}
    \begin{tablenotes}
     \item[$\star$] The redshift is derived based on the \HI\ line centroid 
     \item[$\dagger$] The stellar mass is from SDSS.
    	\end{tablenotes}
    \end{threeparttable}
\end{center}
\end{table*}

\begin{figure*}
\begin{center}
\includegraphics[width=.2\textwidth]{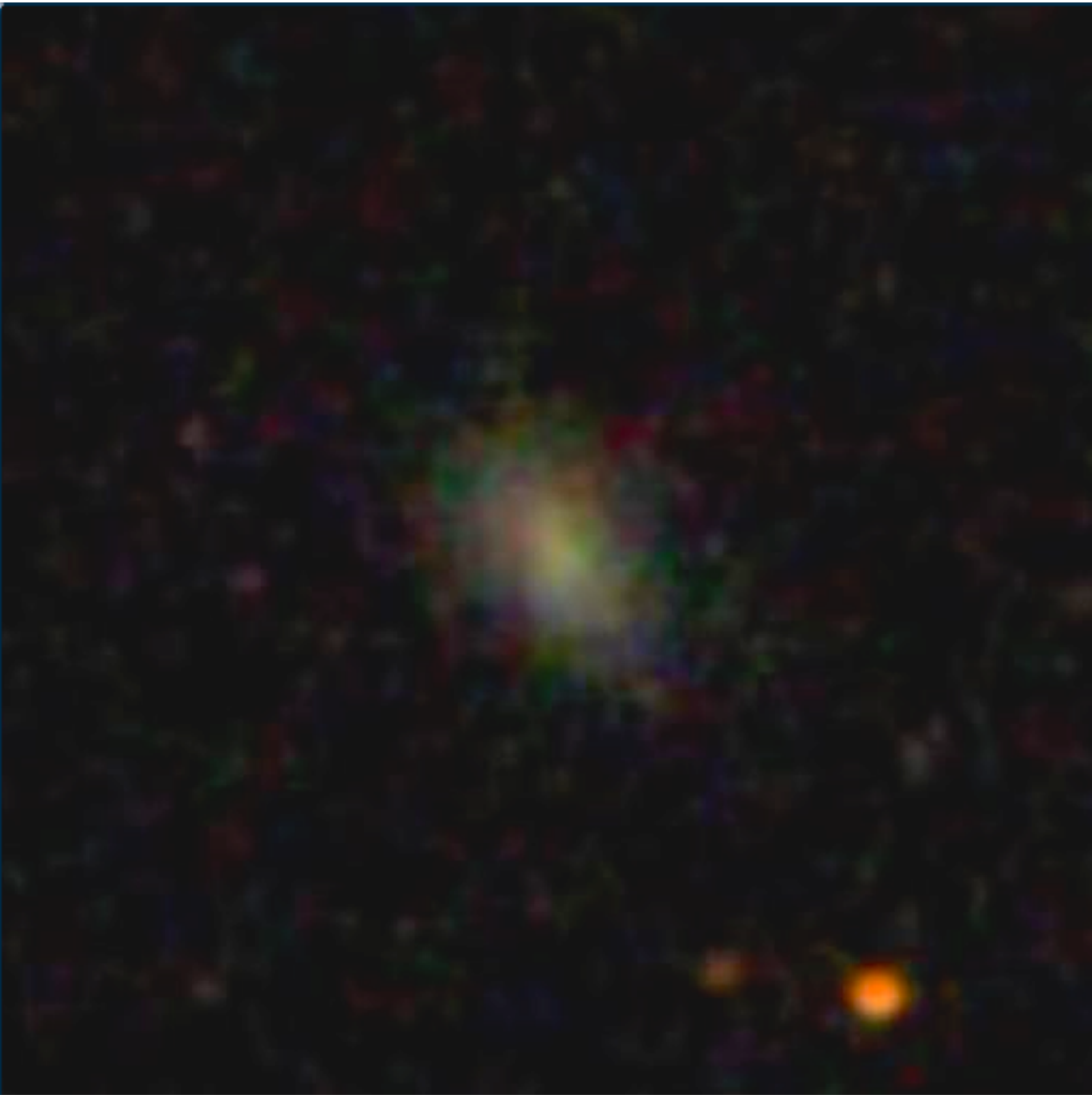}
\includegraphics[width=.35\textwidth]{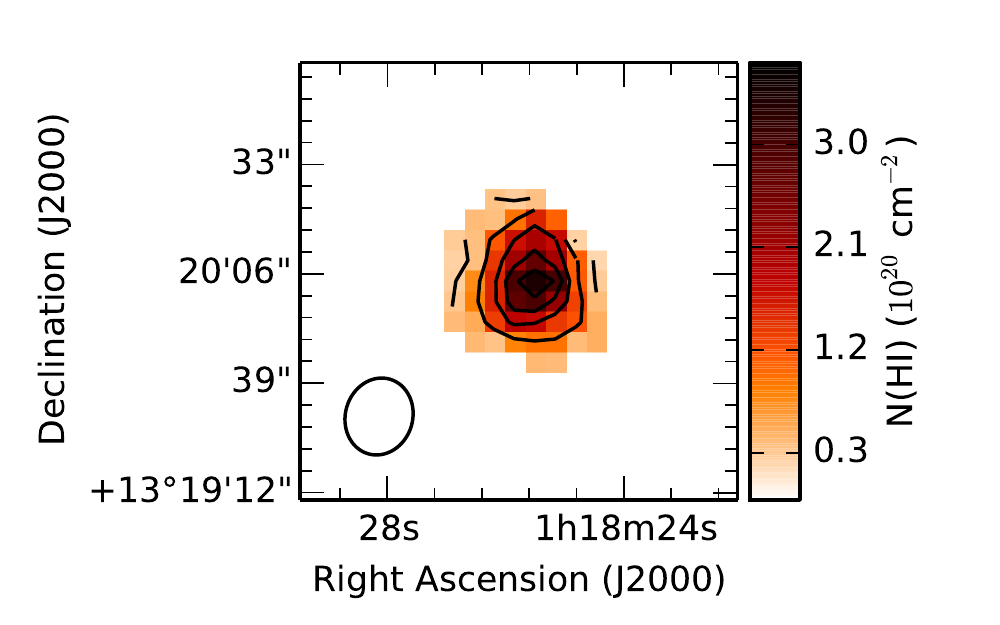}
\includegraphics[width=.35\textwidth]{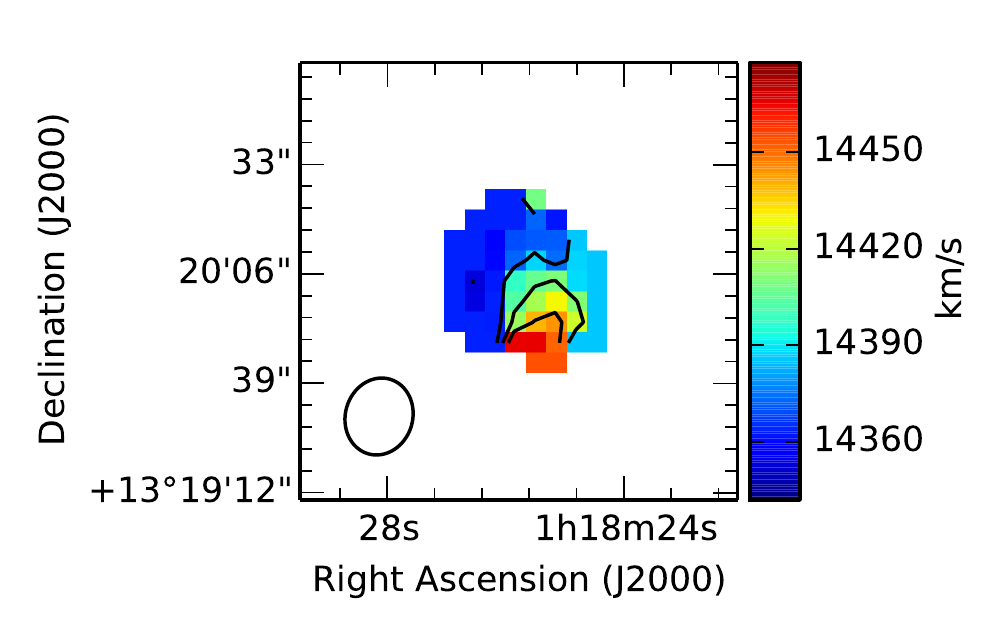}

\includegraphics[width=.2\textwidth]{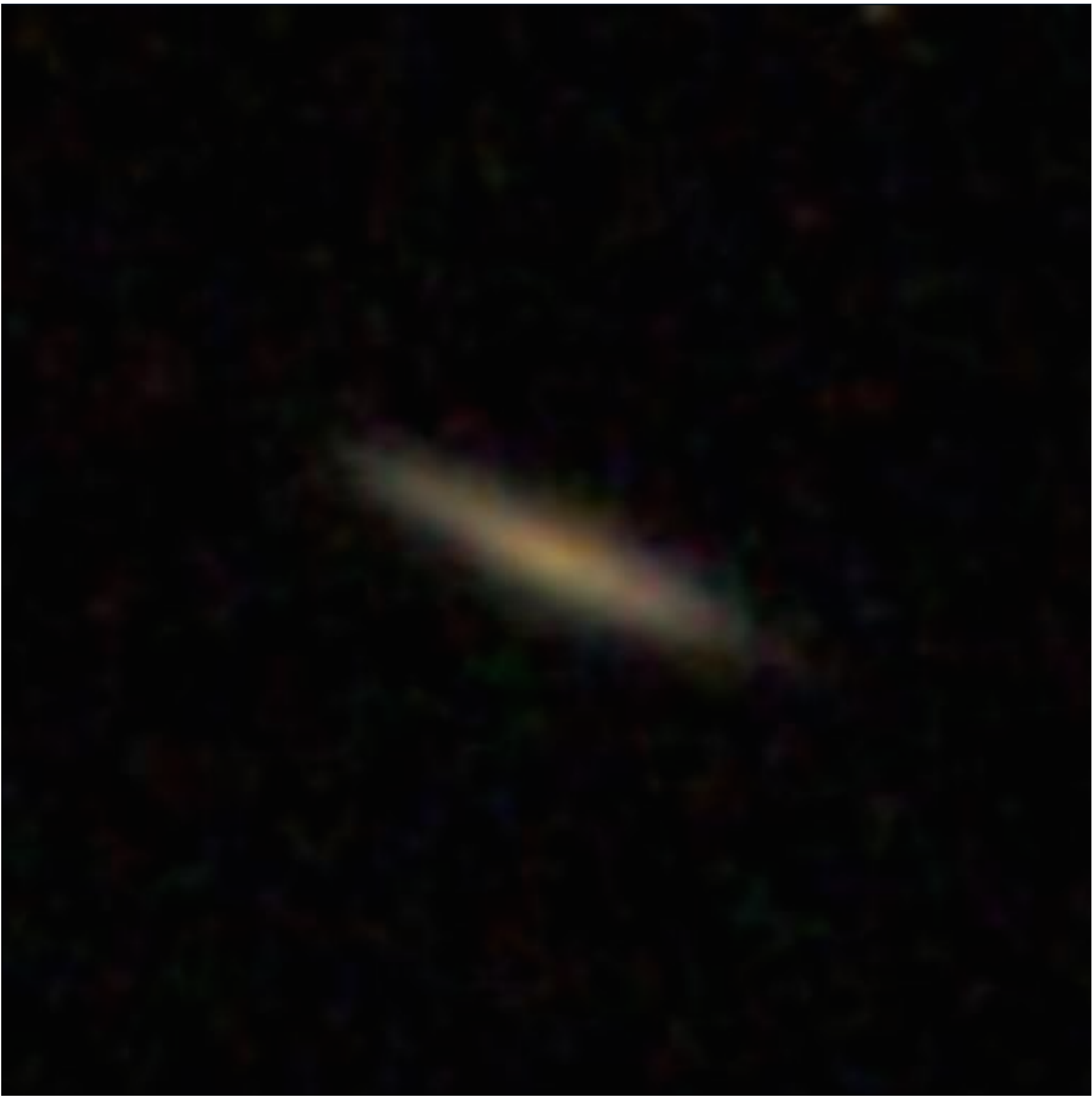}
\includegraphics[width=.35\textwidth]{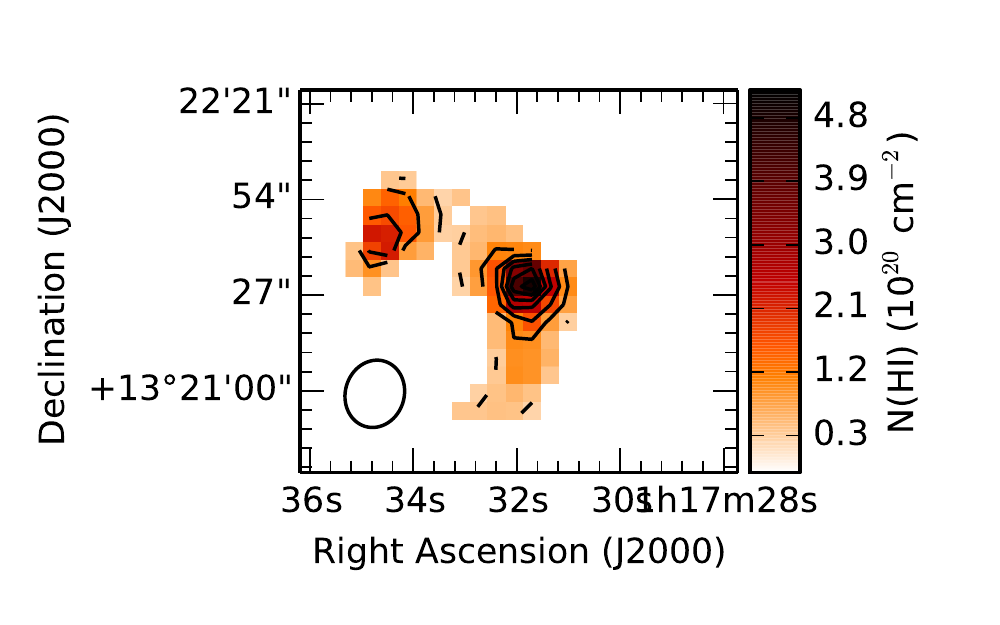}
\includegraphics[width=.35\textwidth]{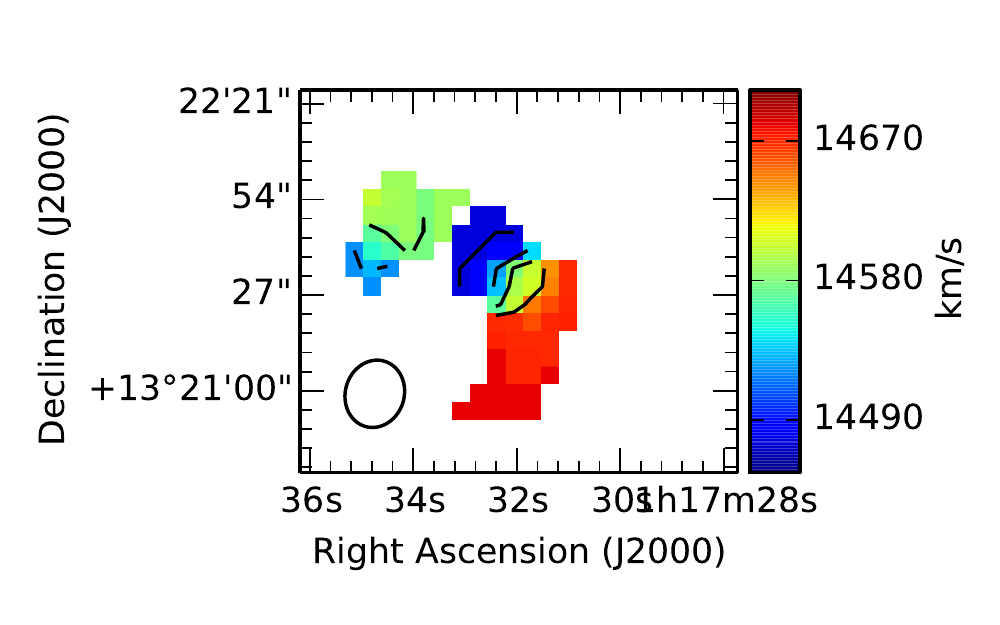}

\includegraphics[width=.2\textwidth]{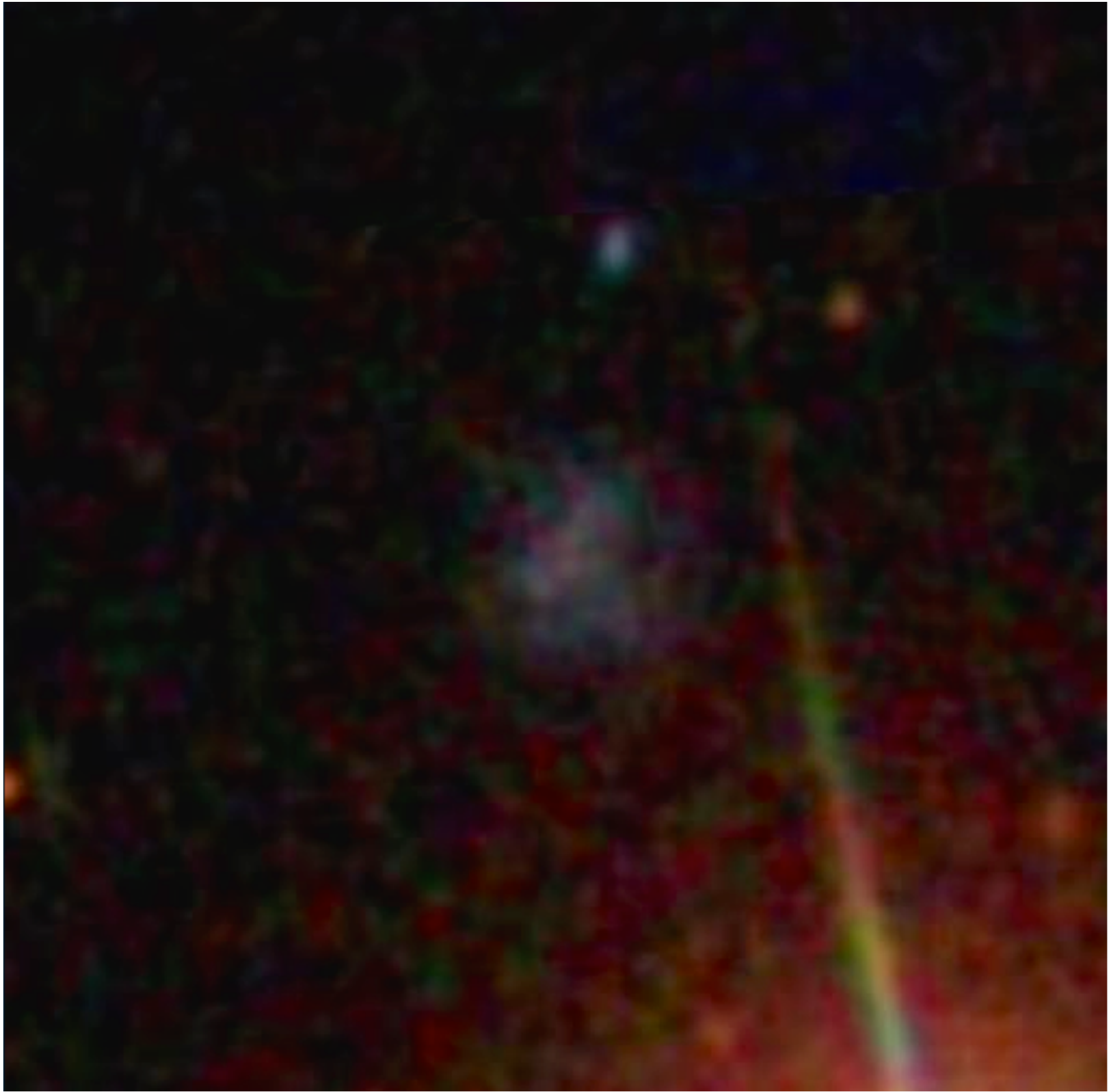}
\includegraphics[width=.35\textwidth]{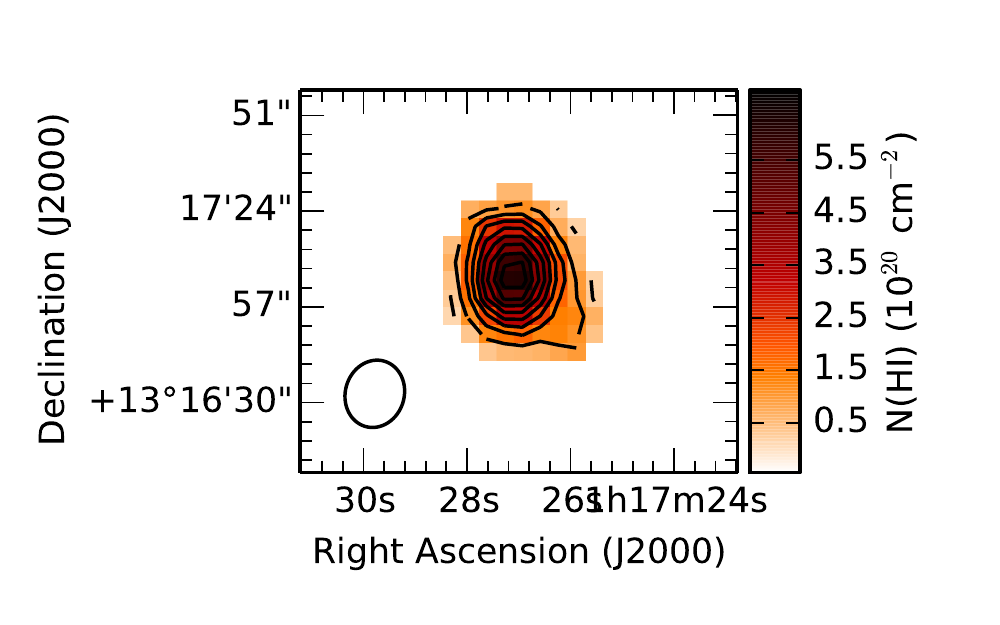}
\includegraphics[width=.35\textwidth]{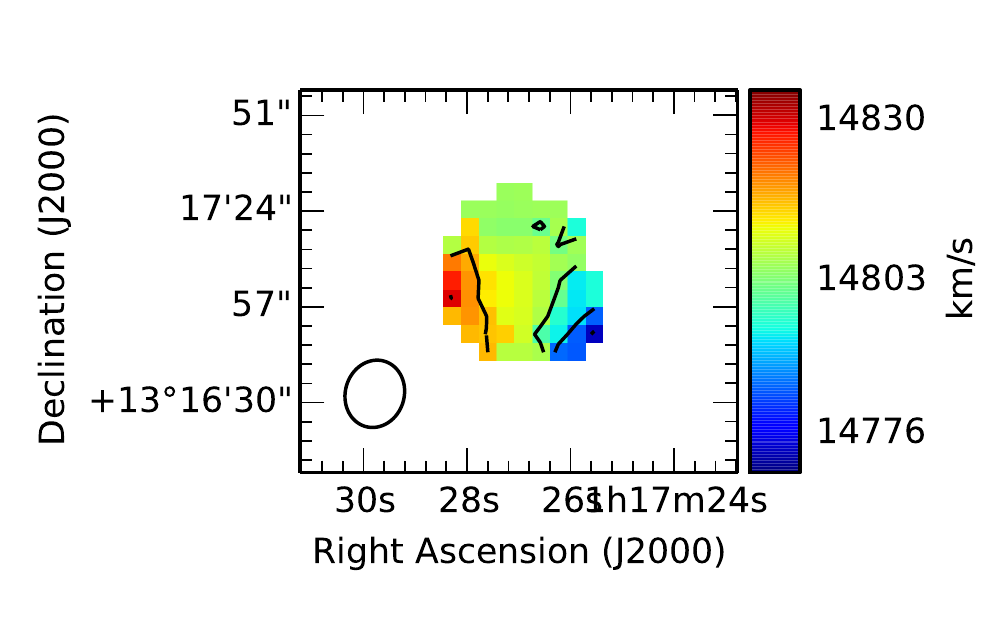}

\includegraphics[width=.2\textwidth]{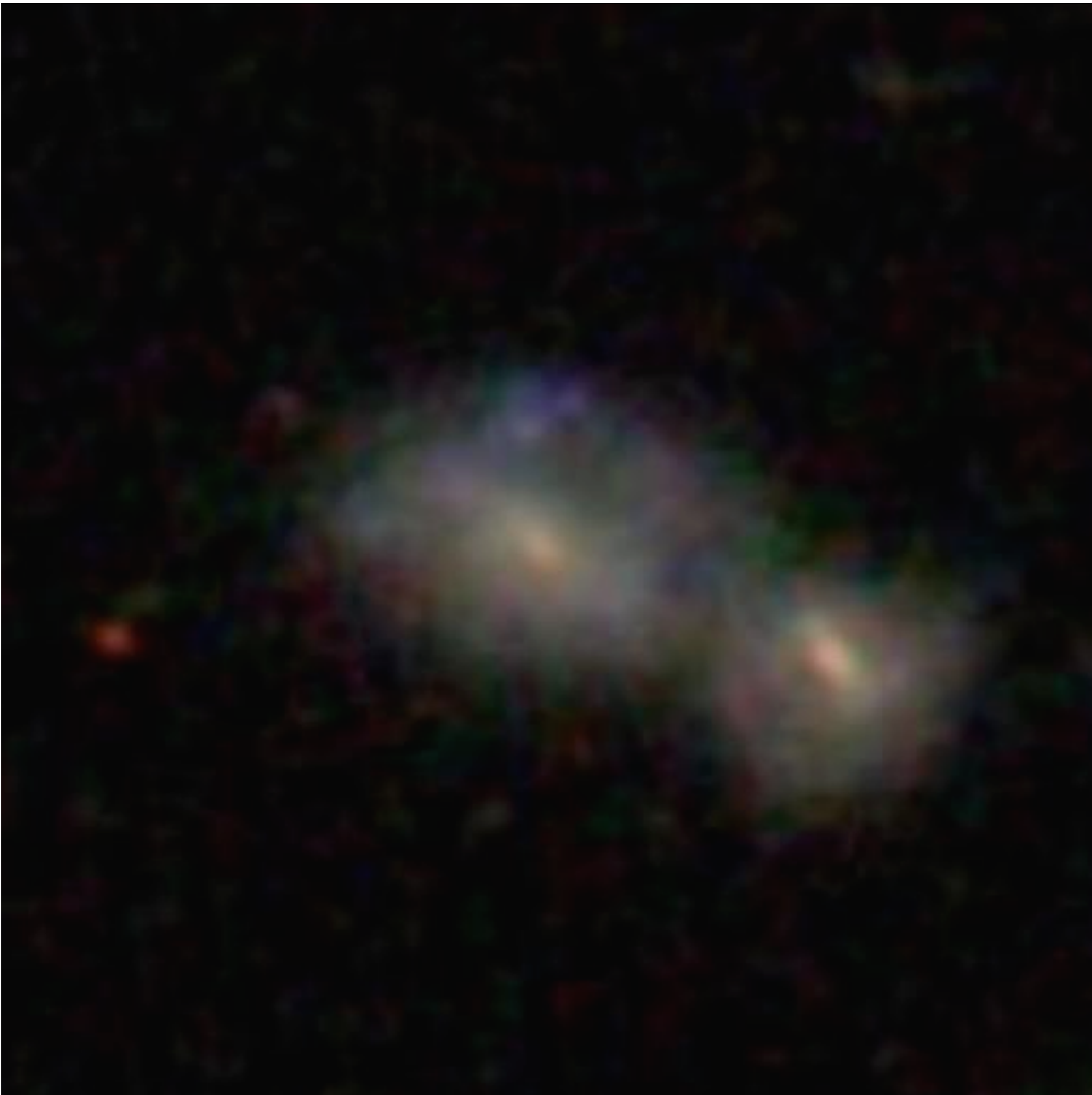}
\includegraphics[width=.35\textwidth]{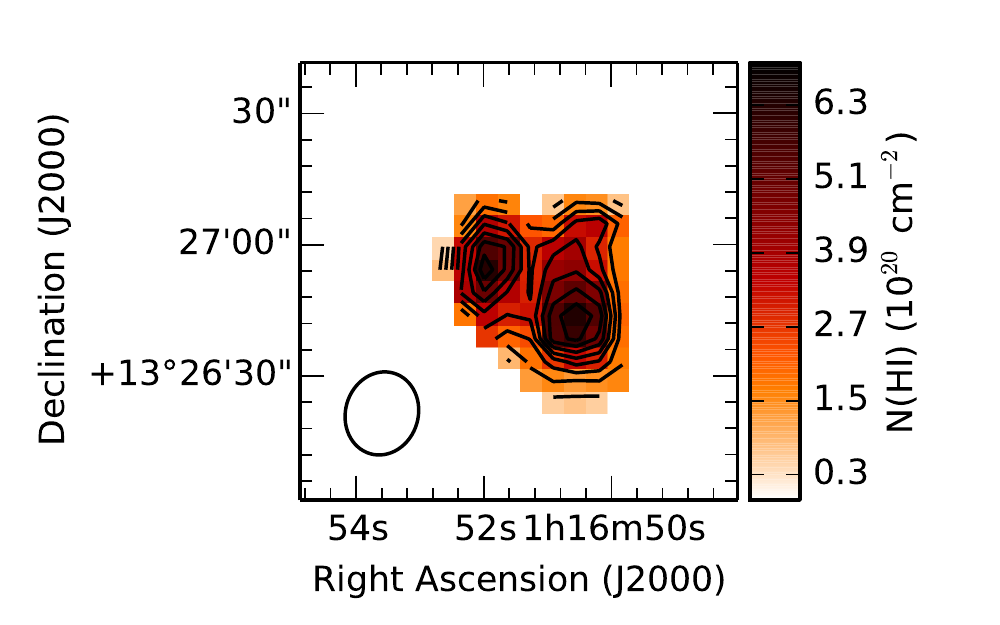}
\includegraphics[width=.35\textwidth]{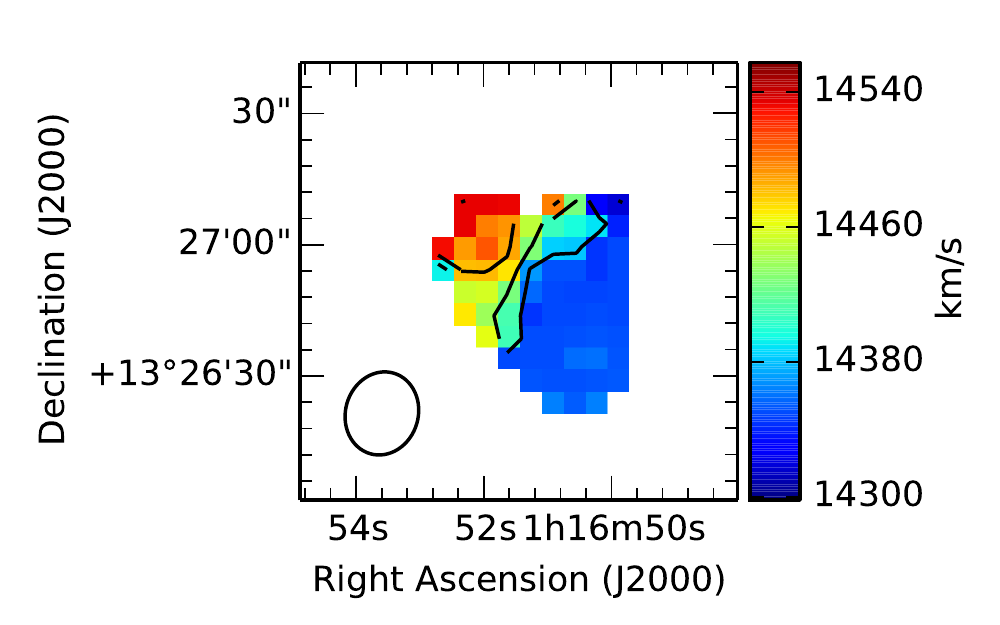}

\caption{SDSS images (left, the size of the images is 50 arcsec), \HI\ column density (middle), and first-moment (velocity field, right) maps of \HI\ detections in the environment of GASS 3505. The \HI\ column density contours range between 3$\sigma$ and peak intensity, with steps of 6$\sigma$. The order of the images from the top to the bottom is: J0118+1320, J0117+1321, J0117+1317, J0116+1327.}\label{fig:HImaps_environemnt}
\end{center}
\end{figure*}

%% file: GASS3505_accepted.bbl
\begin{thebibliography}{Dekel}

\bibitem[\protect\citeauthoryear{Balcells et 
al.}{2001}]{Balcells2001} Balcells M., van Gorkom J.~H., Sancisi R., 
del Burgo C., 2001, AJ, 122, 1758 

\bibitem[\protect\citeauthoryear{Bekki}{2013}]{Bekki2013} Bekki 
K., 2013, MNRAS, 432, 2298 


\bibitem[\protect\citeauthoryear{Bertin et al.}{2002}]{Bertin2002} 
Bertin E., Mellier Y., Radovich M., Missonnier G., Didelon P., Morin B., 
2002, ASPC, 281, 228 


\bibitem[\protect\citeauthoryear{Bigiel et al.}{2010}]{Bigiel2010} 
Bigiel F., Leroy A., Walter F., Blitz L., Brinks E., de Blok W.~J.~G., 
Madore B., 2010, AJ, 140, 1194 

\bibitem[\protect\citeauthoryear{Binney \& Tremaine}{1987}]{Binney1987} Binney J., Tremaine S., 1987, gady.book,  


\bibitem[\protect\citeauthoryear{Binney}{2004}]{Binney2004} Binney 
J., 2004, MNRAS, 347, 1093 

\bibitem[\protect\citeauthoryear{Blanton 
\& Moustakas}{2009}]{Blanton2009} Blanton M.~R., Moustakas J., 2009, ARA\&A, 47, 159 

\bibitem[\protect\citeauthoryear{Behroozi, Wechsler, \& Conroy}{2013}]{Behroozi2013} Behroozi P.~S., Wechsler R.~H., Conroy C., 2013, ApJ, 770, 57 


\bibitem[\protect\citeauthoryear{Booth et al.}{2009}]{Booth2009} 
Booth R.~S., de Blok W.~J.~G., Jonas J.~L., Fanaroff B., 2009, arXiv, 
arXiv:0910.2935 


\bibitem[\protect\citeauthoryear{Brown et al.}{2015}]{Brown2015} 
Brown T., Catinella B., Cortese L., Kilborn V., Haynes M.~P., Giovanelli 
R., 2015, MNRAS, 452, 2479 

\bibitem[\protect\citeauthoryear{Catinella et 
al.}{2008}]{Catinella2008} Catinella B., Haynes M.~P., Giovanelli R., 
Gardner J.~P., Connolly A.~J., 2008, ApJ, 685, L13 


\bibitem[\protect\citeauthoryear{Catinella et 
al.}{2010}]{Catinella} Catinella B., et al., 2010, MNRAS, 403, 
683 

\bibitem[\protect\citeauthoryear{Catinella et 
al.}{2013}]{Catinella2013} Catinella B., et al., 2013, MNRAS, 436, 34 

\bibitem[\protect\citeauthoryear{Catinella 
\& Cortese}{2015}]{Catinella2015} Catinella B., Cortese L., 2015, MNRAS, 446, 3526 

\bibitem[\protect\citeauthoryear{Chabrier}{2003}]{Chabrier2003} 
Chabrier G., 2003, PASP, 115, 763 

\bibitem[\protect\citeauthoryear{Churchill, Steidel, 
\& Kacprzak}{2005}]{Churchill2005} Churchill C., Steidel C., Kacprzak G., 2005, ASPC, 331, 387 


\bibitem[\protect\citeauthoryear{Cortese 
\& Hughes}{2009}]{Cortese2009} Cortese L., Hughes T.~M., 2009, MNRAS, 400, 1225 


\bibitem[\protect\citeauthoryear{Cortese et 
al.}{2011}]{Cortese2011} Cortese L., Catinella B., Boissier S., 
Boselli A., Heinis S., 2011, MNRAS, 415, 1797 


\bibitem[\protect\citeauthoryear{DeBoer et al.}{2009}]{DeBoer2009} 
DeBoer D.~R., et al., 2009, IEEEP, 97, 1507 

\bibitem[\protect\citeauthoryear{D{\'e}nes, Kilborn, 
\& Koribalski}{2014}]{Denes2014} D{\'e}nes H., Kilborn V.~A., Koribalski B.~S., 2014, MNRAS, 444, 667 


\bibitem[\protect\citeauthoryear{Dekel et al.}{2009}]{Dekel2009} Dekel A., et al., 2009, Nature, 457, 451 

\bibitem[\protect\citeauthoryear{Dressler}{1980}]{Dressler1980} 
Dressler A., 1980, ApJ, 236, 351 

\bibitem[\protect\citeauthoryear{van Driel $\&$ van Woerden}{1991}]{Driel1991} van Driel W., van Woerden H., 1991, A\&A, 243, 71 

\bibitem[\protect\citeauthoryear{Duc et al.}{2015}]{Duc2015} 
Duc P.-A., et al., 2015, MNRAS, 446, 120 

\bibitem[\protect\citeauthoryear{Duffy et al.}{2012}]{Duffy2012} 
Duffy A.~R., Meyer M.~J., Staveley-Smith L., Bernyk M., Croton D.~J., 
Koribalski B.~S., Gerstmann D., Westerlund S., 2012, MNRAS, 426, 3385 

\bibitem[\protect\citeauthoryear{Evslin}{2014}]{Evslin2014} Evslin J., 2014, MNRAS, 440, 1225 


\bibitem[\protect\citeauthoryear{Fabello et 
al.}{2011}]{Fabello2011} Fabello S., Catinella B., Giovanelli R., 
Kauffmann G., Haynes M.~P., Heckman T.~M., Schiminovich D., 2011, MNRAS, 
411, 993 

\bibitem[\protect\citeauthoryear{Fernandez et 
al.}{2015}]{Fernandez2015} Fernandez X., van Gorkom J.~H., Momjian 
E., Chiles Team, 2015, AAS, 225, $\#$427.03 

\bibitem[\protect\citeauthoryear{Fraternali et 
al.}{2002}]{Fraternali2002} Fraternali F., van Moorsel G., Sancisi R., 
Oosterloo T., 2002, AJ, 123, 3124 

\bibitem[\protect\citeauthoryear{Gallazzi et al.}{2008}]{Gallazzi2008} Gallazzi A., Brinchmann J., Charlot S., White S.~D.~M., 2008, MNRAS, 383, 1439 


\bibitem[\protect\citeauthoryear{Ger{\'e}b et 
al.}{2015}]{Gereb2015} Ger{\'e}b K., Morganti R., Oosterloo T.~A., Hoppmann L., Staveley-Smith L., 2015, A\&A, 580, A43 

\bibitem[\protect\citeauthoryear{Giovanelli et 
al.}{2005}]{Giovanelli2005} Giovanelli R., et al., 2005, AJ, 130, 2598 

\bibitem[\protect\citeauthoryear{G{\'o}mez et 
al.}{2003}]{Gomez2003} G{\'o}mez P.~L., et al., 2003, ApJ, 584, 
210 


\bibitem[van Gorkom $\&$ Schiminovich(1997)]{Gorkom1997} van Gorkom, J., $\&$ Schiminovich, D.\ 1997, The Nature of Elliptical Galaxies; 2nd Stromlo Symposium, 116, 310 


\bibitem[\protect\citeauthoryear{Hallenbeck et 
al.}{2014}]{Hallenbeck2014} Hallenbeck G., et al., 2014, AJ, 148, 69 

\bibitem[\protect\citeauthoryear{Haynes 
\& Giovanelli}{1984}]{Haynes1984} Haynes M.~P., Giovanelli R., 1984, AJ, 89, 758 



\bibitem[\protect\citeauthoryear{Haynes et al.}{2011}]{Haynes2011} Haynes M.~P., et al., 2011, AJ, 142, 170 


\bibitem[\protect\citeauthoryear{Heald et 
al.}{2011}]{Heald2011} Heald G., et al., 2011, A\&A, 526, A118 


 \bibitem[\protect\citeauthoryear{Hernquist 
\& Quinn}{1988}]{Hernquist1988} Hernquist L., Quinn P.~J., 1988, ApJ, 331, 682 

\bibitem[\protect\citeauthoryear{Huang et al.}{2012}]{Huang2012} 
Huang S., Haynes M.~P., Giovanelli R., Brinchmann J., 2012, ApJ, 756, 113 


\bibitem[\protect\citeauthoryear{Huang et al.}{2014}]{Huang2014} 
Huang S., et al., 2014, ApJ, 793, 40 


\bibitem[\protect\citeauthoryear{van der Hulst et al.}{1987}]{Hulst1987} van der Hulst J.~M., Skillman E.~D., Kennicutt R.~C., Bothun G.~D., 1987, A$\&$A, 177, 63 

\bibitem[\protect\citeauthoryear{van der Hulst $\&$ Sancisi}{1988}]{Hulst1988} van der Hulst T., Sancisi R., 1988, AJ, 95, 1354 

\bibitem[\protect\citeauthoryear{van der Hulst 
\& Sancisi}{2005}]{Hulst2005} van der Hulst J.~M., Sancisi R., 2005, ASPC, 331, 139 


\bibitem[\protect\citeauthoryear{Ibata et al.}{2001}]{Ibata2001} 
Ibata R., Irwin M., Lewis G., Ferguson A.~M.~N., Tanvir N., 2001, Natur, 
412, 49 

\bibitem[\protect\citeauthoryear{Johnston, Sackett, 
\& Bullock}{2001}]{Johnston2001} Johnston K.~V., Sackett P.~D., Bullock J.~S., 2001, ApJ, 557, 137 

\bibitem[\protect\citeauthoryear{Kacprzak 
\& Churchill}{2011}]{Kacprzak2012} Kacprzak G.~G., Churchill C.~W., 2011, ApJ, 743, L34 

\bibitem[\protect\citeauthoryear{Karachentsev \& Makarov}{1999}]{Karachentsev1999} Karachentsev I.~D., Makarov D.~I., 1999, IAUS, 186, 109 


\bibitem[\protect\citeauthoryear{Kennicutt}{1989}]{Kennicutt1989} 
Kennicutt R.~C., Jr., 1989, ApJ, 344, 685 

\bibitem[\protect\citeauthoryear{Kennicutt, Kobulnicky, 
\& Pizagno}{1998}]{Kennicutt1998} Kennicutt R.~C., Kobulnicky H.~A., Pizagno J.~L., 1998, AAS, 30, 1354 

\bibitem[\protect\citeauthoryear{Kere{\v s} et 
al.}{2005}]{Keres2005} Kere{\v s} D., Katz N., Weinberg D.~H., 
Dav{\'e} R., 2005, MNRAS, 363, 2 


\bibitem[\protect\citeauthoryear{Knapp, Turner, 
\& Cunniffe}{1985}]{Knapp1985} Knapp G.~R., Turner E.~L., Cunniffe P.~E., 1985, AJ, 90, 454 

\bibitem[\protect\citeauthoryear{Larson}{1972}]{Larson1972} Larson 
R.~B., 1972, Natur, 236, 21 

\bibitem[\protect\citeauthoryear{Lemonias et 
al.}{2014}]{Lemonias2014} Lemonias J.~J., Schiminovich D., Catinella 
B., Heckman T.~M., Moran S.~M., 2014, ApJ, 790, 27 


\bibitem[\protect\citeauthoryear{Lewis et al.}{2002}]{Lewis2002} 
Lewis I., et al., 2002, MNRAS, 334, 673 


\bibitem[\protect\citeauthoryear{Malin 
\& Hadley}{1997}]{Malin1997} Malin D., Hadley B., 1997, PASA, 14, 52 


\bibitem[\protect\citeauthoryear{Martin et al.}{2005}]{Martin2005} Martin D.~C., et al., 2005, ApJ, 619, L1 


\bibitem[\protect\citeauthoryear{Martin et al.}{2012}]{Martin2012} 
Martin C.~L., Shapley A.~E., Coil A.~L., Kornei K.~A., Bundy K., Weiner 
B.~J., Noeske K.~G., Schiminovich D., 2012, ApJ, 760, 127 

\bibitem[\protect\citeauthoryear{McConnachie et 
al.}{2003}]{McConnachie2003} McConnachie A.~W., Irwin M.~J., Ibata 
R.~A., Ferguson A.~M.~N., Lewis G.~F., Tanvir N., 2003, MNRAS, 343, 1335 


\bibitem[\protect\citeauthoryear{McMullin et al.}{2007}]{McMullin2007} McMullin J.~P., Waters B., Schiebel D., Young W., Golap K., 2007, ASPC, 376, 127 


\bibitem[\protect\citeauthoryear{McQuinn et al.}{2010}]{McQuinn2010} McQuinn K.~B.~W., et al., 2010, ApJ, 724, 49 


\bibitem[\protect\citeauthoryear{Meyer et al.}{2004}]{Meyer2004} Meyer M.~J., et al., 2004, MNRAS, 350, 1195 


\bibitem[\protect\citeauthoryear{Morganti et al.}{1997}]{Morganti1997} Morganti R., Sadler E.~M., Oosterloo T., Pizzella A., Bertola F., 1997, AJ, 113, 937 


\bibitem[\protect\citeauthoryear{Morganti et al.}{2006}]{Morganti2006} Morganti R., et al., 2006, MNRAS, 371, 157 


 \bibitem[\protect\citeauthoryear{Oosterloo, Fraternali, 
\& Sancisi}{2007}]{Oosterloo2007} Oosterloo T., Fraternali F., Sancisi R., 2007, AJ, 134, 1019 

\bibitem[\protect\citeauthoryear{Oosterloo et al.}{2010a}]{Oosterloo2010} Oosterloo T., et al., 2010a, MNRAS, 409, 500 

\bibitem[\protect\citeauthoryear{Oosterloo, Verheijen, 
\& van Cappellen}{2010b}]{Oosterloo2010b} Oosterloo T., Verheijen M., van Cappellen W., 2010b, iska.meet, 43 

\bibitem[\protect\citeauthoryear{Pickering et al.}{1997}]{Pickering1997} Pickering T.~E., Impey C.~D., van Gorkom 
J.~H., Bothun G.~D., 1997, AJ, 114, 1858 


\bibitem[\protect\citeauthoryear{Roberts 
\& Haynes}{1994}]{Roberts1994} Roberts M.~S., Haynes M.~P., 1994, ARA\&A, 32, 115 


\bibitem[\protect\citeauthoryear{Rots et al.}{1990}]{Rots1990} 
Rots A.~H., Bosma A., van der Hulst J.~M., Athanassoula E., Crane P.~C., 
1990, AJ, 100, 387 


\bibitem[\protect\citeauthoryear{Sadler, Oosterloo, 
\& Morganti}{2002}]{Sadler2002} Sadler E.~M., Oosterloo T., Morganti R., 2002, ASPC, 273, 215 

\bibitem[\protect\citeauthoryear{Saintonge et al.}{2011}]{Saintonge2011} Saintonge A., et al., 2011a, MNRAS, 415, 32 


\bibitem[\protect\citeauthoryear{Salpeter}{1955}]{Salpeter1955} Salpeter E.~E., 1955, ApJ, 121, 161 


\bibitem[\protect\citeauthoryear{Sancisi et 
al.}{2008}]{Sancisi2008} Sancisi R., Fraternali F., Oosterloo T., van der Hulst T., 2008, A\&ARv, 15, 189 


\bibitem[\protect\citeauthoryear{Schmidt}{1959}]{Schmidt1959} 
Schmidt M., 1959, ApJ, 129, 243 

\bibitem[\protect\citeauthoryear{Serra et 
al.}{2006}]{Serra2006} Serra P., Trager S.~C., van der Hulst J.~M., Oosterloo T.~A., Morganti R., 2006, A\&A, 453, 493 


\bibitem[\protect\citeauthoryear{Serra et al.}{2012}]{Serra} 
Serra P., et al., 2012, MNRAS, 422, 1835 


\bibitem[\protect\citeauthoryear{Veron-Cetty et 
al.}{1995}]{Veron1995} Veron-Cetty M.-P., Woltjer L., Ekers R.~D., Staveley-Smith L., 1995, A$\&$A, 297, L79 

\bibitem[\protect\citeauthoryear{Wang et al.}{2013}]{Wang2013} 
Wang J., et al., 2013, MNRAS, 433, 270 

\bibitem[\protect\citeauthoryear{Wardle 
\& Knapp}{1985}]{Wardle1985} Wardle M.~J., Knapp G.~R., 1985, BAAS, 17, 611 


\bibitem[\protect\citeauthoryear{Wyder et al.}{2007}]{Wyder2007} Wyder T.~K., et al., 2007, ApJS, 173, 293 


\bibitem[\protect\citeauthoryear{van Woerden, van Driel, 
\& Schwarz}{1983}]{vanWoerden1983} van Woerden H., van Driel W., Schwarz U.~J., 1983, IAUS, 100, 99 

\bibitem[\protect\citeauthoryear{York et al.}{2000}]{York} York D.~G., et al., 2000, AJ, 120, 1579 

\bibitem[\protect\citeauthoryear{Yun, Ho, 
\& Lo}{1994}]{Yun1994} Yun M.~S., Ho P.~T.~P., Lo K.~Y., 1994, Natur, 372, 530 


\bibitem[\protect\citeauthoryear{Zibetti, Charlot, $\&$ Rix}{2009}]{Zibetti2009} Zibetti S., Charlot S., Rix H.-W., 2009, MNRAS, 400, 1181 

\bibitem[\protect\citeauthoryear{Zwaan et al.}{2005}]{Zwaan2005} Zwaan M.~A., Meyer M.~J., Staveley-Smith L., Webster R.~L., 2005, MNRAS, 359, L30 



\end{thebibliography}
